\documentclass[[preprint, 11pt]{article} 
\input{epsf}
\usepackage[utf8]{inputenc}
\usepackage{amsmath}
\usepackage{amssymb}
\input{epsf}
\usepackage{natbib}
\usepackage[margin=1in]{geometry}
\usepackage[raggedright]{titlesec}
\usepackage[hang,flushmargin]{footmisc} 
\usepackage{graphicx}
\usepackage{framed}
\usepackage{color}
\usepackage{yfonts}
\usepackage[dvipsnames]{xcolor}
\usepackage{pifont}
\usepackage{amsmath,amssymb}
\usepackage{multirow}
\usepackage{amsfonts}
\usepackage{subfigure}
\usepackage{caption}
\usepackage{threeparttable}
\usepackage{esvect}
\usepackage{pgf}
\usepackage{tikz}
\usepackage{tikz-cd}
\usepackage{fancyvrb}
\usetikzlibrary{positioning}
\usepackage{bbm}
\usepackage{colortbl}
\usepackage{pdflscape}
\usepackage{booktabs}
\usepackage{amsmath}
\usepackage{mathdots}
\usepackage{yhmath}
\usepackage{cancel}
\usepackage{siunitx}
\usepackage{array}
\usepackage{multirow}
\usepackage{gensymb}
\usepackage{tabularx}
\usepackage{booktabs}
\usetikzlibrary{fadings}
\usetikzlibrary{patterns}
\usetikzlibrary{shadows.blur}
\usetikzlibrary{shapes}

\usepackage{hyperref}
\hypersetup{
    colorlinks=true,
    linkcolor=NavyBlue,
    filecolor=magenta,      
    urlcolor=cyan,
    citecolor=black
}
 
\usepackage{rotating}
\usepackage{ntheorem}
\makeatletter
\renewtheoremstyle{plain} % Adds automatic line break, if heading is too long
{\item{\theorem@headerfont ##1\ ##2\theorem@separator}~}
{\item{\theorem@headerfont ##1\ ##2\ (##3)\theorem@separator}~}
\makeatother

\theoremheaderfont{\normalfont\bfseries}

\newtheorem{theorem}{Theorem}[section]
\newtheorem{definition}{Definition}[section]

\newtheorem{proposition}{Proposition}[section]
\newtheorem{corollary}{Corollary}[section]
\newtheorem{example}{Example}[section]
\newtheorem{assumption}{Assumption}

\newenvironment{proof}{\paragraph{Proof:}}{\hfill$\square$}

\newcommand{\iid}{\stackrel{iid}{\sim}}
\newcommand{\cip}{\stackrel{p}{\rightarrow}}
\newcommand{\cid}{\stackrel{d}{\rightarrow}}
\newcommand{\E}{\mathbb{E}}

\newcommand{\var}{\text{var}}
\newcommand{\cov}{\text{cov}}

\newcommand{\cor}{\text{cor}}

\newcommand{\bX}{\mathbf{X}}

\newcommand{\bU}{\mathbf{U}}

\newcommand{\R}{\mathbbm{R}}

\newcommand{\indep}{\raisebox{0.05em}{\rotatebox[origin=c]{90}{$\models$}}}%Perpendicular symbol
\definecolor{shadecolor}{gray}{0.9}

\newcommand{\ubrace}[2]{{\underbrace{#1}_{#2}}}
\newsavebox{\usqrtbox}
\newcommand{\usqrt}[1]{% square root with underbraced material
  \sbox{\usqrtbox}{%
    \renewcommand{\ubrace}[2]{##1}% deliver only the main part
    $\displaystyle#1$%
  }%
  \sqrt{\phantom{\usebox{\usqrtbox}}}\hspace*{-\wd\usqrtbox}#1%
}

\tikzset{every picture/.style={line width=0.75pt}} %set default line width to 0.75pt        

\usepackage{enumitem}
\newlist{Step}{enumerate}{2}
\setlist[Step]{label={{Step \arabic*.}}, leftmargin=*}

\usepackage{setspace}
 \makeatletter
\newcommand\circled[1]{%
  \mathpalette\@circled{#1}%
}
\newcommand\@circled[2]{%
  \tikz[baseline=(math.base)] \node[draw,circle,inner sep=2pt] (math) {$\m@th#1#2$};%
}
\makeatother

 \makeatletter
\newcommand\circledblue[1]{%
  \mathpalette\@circledblue{#1}%
}
\newcommand\@circledblue[2]{%
  \tikz[baseline=(math.base)] \node[draw,circle, fill=blue!20, inner sep=2pt] (math) {$\m@th#1#2$};%
 }

\renewenvironment{abstract}
 {\begin{center}\normalsize\textsc{Abstract}%
 \end{center}\begin{quote}\normalsize}
 {\end{quote}}

\title{Variance-based sensitivity analysis for weighting estimators results in more informative bounds\footnote{The authors would like to thank Erin Hartman, Kevin Guo, Alex Franks, Avi Feller, Peng Ding, Nicole Pashley, and Cyrus Samii for helpful comments. Melody Huang is supported by the National Science Foundation Graduate Research Fellowship under Grant No. 2146752, and Samuel D. Pimentel is supported by the National Science Foundation under Grant No. 2142146. Any opinion, findings, and conclusions or recommendations expressed in this material are those of the authors(s) and do not necessarily reflect the views of the National Science Foundation.}}
\author{Melody Huang and Samuel D. Pimentel}
\date{\today}

\begin{document}
\maketitle 
\begin{abstract}
Weighting methods are popular tools for estimating causal effects; assessing their robustness under unobserved confounding is important in practice. In the following paper, we introduce a new set of sensitivity models called the ``variance-based sensitivity model’’. The variance-based sensitivity model characterizes the bias from omitting a confounder by bounding distributional differences that arise in the weights from omitting a confounder, with several notable innovations over existing approaches. First, the variance-based sensitivity model can be parameterized by an $R^2$ parameter that is both standardized and bounded. We introduce a formal benchmarking procedure that allows researchers to use observed covariates to reason about plausible parameter values in an interpretable and transparent way. Second, we show that researchers can estimate valid confidence intervals under the variance-based sensitivity model, and provide extensions for incorporating substantive knowledge about the confounder to help tighten the intervals. Last, we demonstrate, both empirically and theoretically, that the variance-based sensitivity model can provide improvements on both the stability and tightness of the estimated confidence intervals over existing methods. We illustrate our proposed approach on a study examining blood mercury levels using the National Health and Nutrition Examination Survey (NHANES).  

\end{abstract}

\clearpage 
\doublespacing

\section{Introduction}
In observational studies of causal effects, researchers must address possible confounding effects from non-random treatment assignment.  Typically, one relies on pre-treatment covariates either to re-weight units based on propensity of treatment, or to model the outcome of interest. In practice, researchers have no way of knowing whether the included covariates fully capture the confounding effects. When confounders are omitted, the resulting estimates will be biased. Sensitivity analyses speak to this concern by allowing researchers to assess the robustness of their point estimates to omitted confounders. In a sensitivity analysis, a researcher introduces a parameter describing the amount of unobserved confounding present and redoes the analysis under different values of this parameter, determining the set of values for which the results of the study will be reversed.  The robustness of the study may then be evaluated by reasoning about the plausibility of these values. 

In contrast to typical estimands, parameters in sensitivity analysis are inherently unindentifiable, because they are designed to describe an omitted variable. Thus, there exists a trade-off between how complex the sensitivity analysis is, and how informative the sensitivity analysis can be. For example, \cite{dahabreh2019sensitivity} proposed a sensitivity analysis in which researchers can obtain both an adjusted point estimate and the associated uncertainty from omitting a confounder. However, the sensitivity analysis requires researchers to directly model the bias that arises from omitting a confounder. In contrast, \cite{zhao2019sensitivity} introduced a sensitivity analysis that only requires one parameter and allows researchers to estimate confidence intervals that account for the unobserved confounder. However, the resulting intervals are often extremely wide, making it difficult to reason about whether or not there is sensitivity from omitting a confounder.

In the following paper, we introduce a new set of sensitivity models known as the \textit{variance-based sensitivity model} that provide the flexibility and generality of existing sensitivity analyses, while simultaneously allowing researchers to estimate narrower, more informative bounds for weighted estimators. The proposed sensitivity models constrain distributional differences in the weights that arise from omitting a confounder, and do not rely on additional assumptions on the outcome, confounder, or treatment assignment mechanism.

The paper provides three primary contributions. First, the proposed variance-based sensitivity model can be parameterized by a single sensitivity parameter, an $R^2$ measure. Unlike previously proposed sensitivity analyses, our parameter is both bounded and standardized on an interval of 0 to 1. We develop formal benchmarking approaches that allow researchers to use observed covariates to reason about plausible values for the $R^2$ value, providing much-needed interpretability.

Second, we introduce a method for estimation of valid confidence intervals under the variance-based sensitivity model. We give a closed-form solution for the maximum bias that can occur for a fixed set of sensitivity models, which we denote the \textit{optimal bias bound}. We also provide extensions for incorporating additional substantive knowledge about the confounder into the optimal bias bound, further restricting the range of plausible bias. 

Finally, we show that a variance-based sensitivity analysis can be formulated as a bias maximization problem, with a constraint on the weighted average error. We formalize the relationship between the variance-based sensitivity model and alternative approaches, which rely on constraining a worst-case error. By moving away from characterizing bias from the perspective of a worst-case error, variance-based sensitivity analysis is able to estimate more informative and stable bounds. 

The paper is organized as follows. Section \ref{sec:background} gives set-up and notation. Section \ref{sec:vbm} introduces the variance-based sensitivity model. Section \ref{sec:msm_compare}, compares the proposed sensitivity models to alternative sensitivity approaches. Proofs and extended discussion are provided in the Appendix. 

\section{Background} \label{sec:background} 
\subsection{Set-Up and Notation}
To begin, we consider an observational study with $n$ individuals. 
Define $Z_i$ as a binary treatment assignment variable, where $Z_i= 1$ when unit $i$ is assigned to treatment, and 0 otherwise, $Y_i(1)$ and $Y_i(0)$ are potential outcomes, and  $\mathcal{X}_i$ is a vector of pre-treatment covariates. Let the tuple $(Y_i(1), Y_i(0), X_i, Z_i)$ for all $i \in \{1, …, n\}$  be independently and identically distributed from an arbitrary joint distribution, where $Y(1), Y(0) \in \mathbbm{R}^n$, $Z_i \in \{0,1\}^n$ and $\mathcal{X} \in \R^{n\times p}$. 

Throughout, we invoke the standard SUTVA assumption---i.e., no interference, with treatments identically administered across all units \citep{rubin1980SUTVA}. Thus observed outcomes $Y_i$ can be written as $Y_i := Y_i(1) \cdot Z_i + Y_i(0) \cdot (1-Z_i)$. 
Because treatments are not randomly assigned in an observational study, we must also invoke an additional assumption, known as the conditional ignorability of treatment assignment: 

\begin{assumption}[Conditional Ignorability of Treatment Assignment]\label{assump:cond_ignor}
$$Y_i(1), Y_i(0) \ \indep \ Z_i \mid \mathcal{X}_i$$
\end{assumption} 
Assumption \ref{assump:cond_ignor} states that conditional on a set of pre-treatment covariates $\mathcal{X}$, treatment assignment is independent of potential outcomes (i.e. no further confounding remains).

In addition to Assumption \ref{assump:cond_ignor}, we also assume overlap, such that conditional on some set of pre-treatment covariates, the probability of being assigned treatment is non-zero \citep{rosenbaum1983assessing}: 
\begin{assumption}[Overlap] 
For all $x \in \mathcal{X}$, $0 < P(Z_i = 1 \mid \mathcal{X} = x) < 1$.
\end{assumption} 

Our primary estimand is the average treatment effect for the treated (ATT): 
$$\tau := \E(Y_i(1) - Y_i(0) \mid Z_i = 1).$$
However, the proposed method can easily be extended for estimating the average treatment effect (ATE). Furthermore, all proofs and derivations are done with respect to a general missingness indicator, such that the results can be applied to general, missing data settings, such as weighting for external validity or survey weighting (see Appendix \ref{app:missingness} for more discussion and details). 

Weighted estimators, which are popular for estimating causal effects in observational setting, adjust for distributional differences in the pre-treatment covariates $\mathcal{X}$ across the treatment and control groups: 
$$\hat \tau_W =  \frac{1}{\sum_{i = 1}^n Z_i} \sum_{i=1}^n Z_i Y_i -
\underbrace{\frac{\sum_{i=1}^n (1-Z_i) Y_i w_i}{\sum_{i=1}^n (1-Z_i) w_i}}_{\text{Weighted Control Mean}}.$$
A common choice of weights is inverse propensity weights, where $w_i = P(Z_i = 1 \mid \mathcal{X})/P(Z_i = 0 \mid \mathcal{X})$. These weights are often constructed using a logistic regression to predict the probability of treatment assignment. Recent balancing approaches provide a semi-parametric option for researchers to estimate weights by minimizing the distributional difference between the treatment and control groups, without modeling the underlying probabilities (see \cite{ben2020balancing} for a recent review on balancing weights). 

Under the correct specification of the weights, if the correct set of covariates are included, the weighted estimator will be a consistent and unbiased estimate of the true treatment effect. However, in practice, there is no way of knowing whether Assumption \ref{assump:cond_ignor} holds. We propose a set of sensitivity models that characterize the bias of a weighted estimator when researchers omit a variable from the set $\mathcal{X}$. More specifically, we will assume that $\mathcal{X} = \{\bX, \bU\}$, such that both $\bX$ and $\bU$ are necessary for Assumption \ref{assump:cond_ignor} to hold. We assume that researchers have otherwise correctly specified the weights.\footnote{For example, the framework does not explicitly account for cases in which researchers are using a probit model, when the true underlying data generating process is logistic. However, mis-specification concerns can also be addressed with the proposed framework, if researchers can write the mis-specification error as an omitted variable problem. A simple example of this is if a linear probability model is used, $\bU$ can include non-linear functions of $\bX$ that matter for modeling selection. We provide more discussion in Appendix \ref{app:phiX}.} We define $w$ as the weights that include only $\bX$, and $w^*$ as the ideal weights that include both $\bX$ and $\bU$. We refer to the omitted variables, $\bU$, as \textit{confounders}. We note that the sensitivity framework will not explicitly account for settings in which (1) researchers estimate uniform weights (i.e., no adjustment for confounding), or (2) the true ideal weights are uniform. Finally, we will assume, without loss of generality, that both $w$ and $w^*$ are centered at mean 1. 

\subsection{Related Literature}
A popular approach for assessing the robustness of weighted estimates to omitted confounders uses the marginal sensitivity model \citep{tan2006distributional}, in which researchers posit a bound, $\Lambda$, on the individual-level error in the weights that can arise under unobserved confounding:
$$\Lambda^{-1} \leq \frac{w_i^*}{w_i} \leq \Lambda, \ \ \ \ \text{ for } i = 1, ..., n,$$
where $\Lambda \geq 1$. $\Lambda$ represents the largest possible error that can arise from omitting a confounder. Researchers can bound the maximum and minimum bias that arises under a fixed $\Lambda$, and use a percentile bootstrap to estimate valid confidence intervals \citep{zhao2019sensitivity}. 

In practice, the true $\Lambda$ is unknown, so to conduct the sensitivity analysis, researchers posit increasing values of $\Lambda$ until the estimated confidence intervals contain zero. The minimum $\Lambda$ value for which the estimated intervals cross zero is denoted as $\Lambda^*$. If $\Lambda^*$ is close to 1, even a small amount of error from omitting a confounder could result in an estimated effect becoming insignificant. On the other hand, if $\Lambda^*$ is much larger than 1, estimated effects are only sensitive to very strong unmeasured confounders.

While the marginal sensitivity model guarantees valid intervals asymptotically, in practice, the intervals tend to be extremely conservative. This means that the estimated intervals often include the null estimate, even under low amounts of confounding. As such, when researchers' estimated bounds imply an estimated effect is no longer statistically significant, it is difficult to distinguish if this is a sign that there is sensitivity to an omitted confounder, or if the sensitivity model is overly pessimistic. Tightening these intervals often requires researchers to invoke additional constraints in the models, or parametrically model the outcomes in some way \citep{dorn2021sharp, nie2021covariate}. Furthermore, the underlying sensitivity parameter in the marginal sensitivity model is dependent on the worst-case error that arises from omitting a confounder. This is inherently difficult to reason about in practice, as the true value of the parameter will depend on outliers and, in asymptotic settings, can be infinitely large. 

We now propose a new set of sensitivity models, the \textit{variance-based sensitivity model}. The proposed framework can be viewed as a one parameter generalization of existing sensitivity frameworks in the literature that rely on bounding the error in the weights from omitting a confounder (i.e., \cite{hartman2022survey}, \cite{hong2021did}, \cite{shen2011sensitivity}). We provide several key contributions. First, unlike the frameworks proposed by \cite{hong2021did} and \cite{shen2011sensitivity}, the variance-based sensitivity model introduce a standardized and bounded parameterization of the confounding strength, which can help improve transparency and interpretability for applied researchers. Second, the aforementioned sensitivity analyses do not engage with how potential confounders may affect their inference, and are limited to discussions about movements in the point estimate. In contrast, the variance-based sensitivity model provides a method for researchers to estimate valid asymptotic confidence intervals for fixed level of confounding. 

Furthermore, we formalize the connection between these variance-based approaches to alternative sensitivity approaches, which formulate sensitivity models as optimization problems. In particular, we demonstrate that the variance-based sensitivity model can be viewed as a constrained weighted $L_2$ norm problem, which provides a framework to compare the proposed sensitivity models with the marginal sensitivity model. Moving away from a worst-case error parameterization of the error allows researchers to obtain more informative and stable bounds under the variance-based sensitivity model. The benefits of constraining a weighted $L_2$ norm instead of a worst-case error is conceptually similar to the advantages highlighted in \cite{zhang2022bounds}, in which authors consider a constraint on $L_2$ norms, and \cite{kallus2018confounding}, which introduces an $L_1$ norm, with the added benefit of having an interpretable sensitivity parameter in the form of an $R^2$ value. 

\subsection{Running Example: NHANES}
Throughout the paper, we perform a re-analysis of a study presented in \cite{zhao2018cross} (as well as \cite{zhao2019sensitivity} and \cite{soriano2021interpretable}), analyzing the effects of fish consumption on blood mercury levels. More specifically, we use data from the 2013-2014 National Health and Nutrition Examination Survey (NHANES). 

Following the original study, we define the outcome of interest as the total blood mercury (in $\log_2$), measured in micrograms per liter. As such, an estimated treated-control outcome difference of 1 implies that a treated person’s total blood mercury is twice that of an individual in control’s total blood mercury. The treatment is defined by whether or not individuals consumed more than 12 servings of fish or shellfish in the preceding month. There are 234 total treated units and 873 control units. To account for the non-random treatment assignment, we use the available demographic data for the individuals in the survey, which include variables like gender, age, income, race, education, and smoking history to estimate entropy balancing weights \citep{hainmueller2012entropy}. 

The unweighted estimate is 2.37; after accounting for pre-treatment covariates, we obtain a weighted ATT estimate of 2.15 (see Table \ref{tbl:fish_summary} for a summary). Therefore, from our estimate, we expect that on average, a treated individual who consumes more fish will have around 4 times as much total blood mercury than an individual in control. 

\begin{table}[ht] 
\centering 
\begin{tabular}{lccc} \toprule 
& Unweighted (DiM) & IPW \\ \midrule 
Estimated Effect (ATT) & 2.37 (0.10) & 2.15 (0.11)\\ \bottomrule 
\end{tabular} 
\caption{Estimated effect of fish consumption on blood mercury levels. Standard errors are reported in parentheses.} 
\label{tbl:fish_summary} 
\end{table}

\section{The Variance-Based Sensitivity Model} \label{sec:vbm} 
We now introduce the \textit{variance-based sensitivity model}. Section \ref{subsec:defining_vbm} defines the sensitivity models and shows that the set of sensitivity models lends itself naturally to an $R^2$ parameterization. Section \ref{subsec:bias} shows that there exists a closed-form solution for the maximum bias under the variance-based sensitivity model that can be directly estimated from the data. Section \ref{subsec:valid_ints} lays out a method to estimate asymptotically valid confidence intervals under the proposed set of sensitivity models. Sections \ref{subsec:benchmarking} and \ref{subsec:vbm_nhanes} provide formal benchmarking tools to help researchers conduct their sensitivity analyses, and illustrate the proposed approach on the running example respectively. Appendix \ref{sec:relax_rho} provides an extension for researchers to impose constraints on the strength of the relationship between the confounder and the outcome within the variance-based sensitivity model.

\subsection{Defining a New Sensitivity Model} \label{subsec:defining_vbm} 
Instead of constraining the worst-case, individual-level multiplicative error across the weights, we constrain the difference in the distribution of the estimated weights and the distribution of the true weights. We refer to this set as the ``variance-based sensitivity model.''   

\begin{definition}[Variance-Based Sensitivity Model] \label{def:vbm} \mbox{}\\
Let $R^2$ be the residual variation in the true weights $w^*$, not explained by $w$: 
\begin{equation*} 
R^2 := 1 - \underbrace{\frac{\var(w_i \mid Z_i = 0)}{\var(w^*_i \mid Z_i = 0)}}_{{\substack{{\scriptsize{\text{Variation in } w^*}}\\{\scriptsize \text{explained by } w}}}}
\end{equation*} 
Then, for a fixed $R^2 \in [0,1)$, we define the variance-based sensitivity model $\sigma(R^2)$: 
$$\sigma(R^2) \equiv \left\{  w^*_i \in \R^n: 1 \leq \frac{\var(w^*_i \mid Z_i = 0)}{\var(w_i\mid Z_i = 0)} \leq \frac{1}{1-R^2} \right\}.$$
\end{definition} 

The variance-based sensitivity model constrains how different the true weights $w^*$ can be from the estimated weights $w$. This serves as a constraint on the residual imbalance in the omitted variable. More formally, we decompose the true weight $w^*$ into two components: (1) the weight $w$, and (2) the residual imbalance in $\bU$: 
\begin{align} 
w^* &= \frac{P(Z = 1 \mid \bX, \bU)}{1-P(Z=1 \mid \bX, \bU)} = \underbrace{\frac{P(Z = 1 \mid \bX)}{1-P(Z =1 \mid \bX)}}_{\substack{\text{Population Propensity}\\{\text{Weights } (w)}}} \cdot \underbrace{\frac{P(\bU \mid \bX, Z = 1 )}{P(\bU \mid \bX, Z = 0)}}_{\text{Imbalance in $\bU$}},
\label{eqn:w_decomp} 
\end{align} 
where the imbalance term is a ratio of the conditional probability density function of the omitted variable across the treatment and control groups. The distributional difference between the weights $w$ and the ideal weights $w^*$ will be driven by the imbalance term. Intuitively, if the omitted variable $\bU$ is very imbalanced, then accounting for the omitted variable results in very different values for $w^*$ and $w$. Alternatively, if $\bU$ is not very imbalanced, then including it will result in weights $w^*$ that are very similar to $w$. Limiting the distributional difference between the true weights and estimated weights effectively restricts the amount of residual imbalance in the omitted confounder. In Section \ref{sec:msm_compare}, we show that this is equivalent to constraining a weighted $L_2$ norm of the errors $w^*_i/w_i$, in contrast with the marginal sensitivity model, which constrains an $L_\infty$ norm. 

The distributional difference between the estimated weights and the true weights can be written as a function of an $R^2$ parameter. The $R^2$ parameter represents the residual variation in the true weights, not explained by the estimated weights, and is naturally bounded on an interval of $[0, 1]$. 

\subsection{Optimal Bias Bounds} \label{subsec:bias} 
Valid confidence intervals for a set of sensitivity models must account for two factors: (1) the bias that arises from omitting a confounder, and (2) the uncertainty associated with estimation. In the following subsection, we introduce optimal bias bounds that researchers can estimate for the variance-based sensitivity model, under a fixed $R^2$ value. Section \ref{subsec:valid_ints} introduces a percentile bootstrap approach for researchers to simultaneously account for uncertainty in estimation.

In the following theorem, we show that for a fixed $R^2$, we can estimate the possible range of bias values. We refer to the minimum and maximum values of these potential bias values as the \textit{optimal bias bounds}. However, unlike the marginal sensitivity models, in which researchers must solve a linear programming problem to identify the extrema, %an advantage to 
the variance-based sensitivity model admits a closed-form solution for the optimal bias. More specifically, the optimal bias bounds are a function of three different components: (1) a correlation bound, which represents the maximum correlation an omitted confounder can have with the outcome of interest; (2) the imbalance (represented by the $R^2$); and (3) a scaling factor.

\begin{theorem}[Optimal Bias Bounds]\label{thm:optim_bounds} \mbox{}\\ 
For a fixed $R^2 \in [0,1)$, the maximum bias under $\sigma(R^2)$ (denoted as $\max_{\tilde w \in \sigma(R^2)} \text{Bias}(\hat \tau_W \mid \tilde w)$) can be written as a function of the following components: 
\begin{align} 
\max_{\tilde w \in \sigma(R^2)} \text{ Bias}(\hat \tau_W \mid \tilde w)
&= \underbrace{\sqrt{1-\cor(w_i, Y_i \mid Z_i = 0)^2}}_{\text{(a) Correlation Bound}} \usqrt{\ubrace{\frac{R^2}{1-R^2}}{\makebox[0pt]{\scriptsize\text{(b) Imbalance}}} \cdot \ubrace{\var(Y_i|Z_i = 0) \cdot \var(w_i | Z_i = 0)}{\makebox[0pt]\scriptsize\text{(c) Scaling Factor}}},
\label{eqn:optim_bounds} 
\end{align} 
with the minimum bias given as the negative of Equation \eqref{eqn:optim_bounds}. The optimal bias bounds are given by the minimum and maximum biases.
\end{theorem} 
Theorem \ref{thm:optim_bounds} highlights the different components that affect the magnitude of the bias bounds. We provide more details about each component below. 

\paragraph{Correlation Bound.} The correlation bound, given by Equation \eqref{eqn:optim_bounds}-(a), represents the maximum correlation between imbalance in an omitted confounder and outcome. Intuitively, this is similar to the marginal sensitivity model, in which \cite{dorn2021sharp} demonstrated that the optimal bias bounds are obtained when the imbalance from the omitted confounder is maximally correlated with the outcome. From Equation \eqref{eqn:optim_bounds}-(a), we see that the bound is a function of the correlation between the estimated weights and the outcome. If the estimated weights are highly correlated with the outcome, then the degree to which the residual imbalance in the omitted confounder can be correlated to the outcome is limited, and this bound is lower. However, if the correlation between the estimated weights and the outcome is relatively low, then the possible correlation between omitted-confounder imbalance and outcome has a much larger range. In the worst case, the estimated weights and the outcome are not correlated at all (i.e., $\cor(w_i, Y_i \mid Z_i = 0) = 0$); then, the correlation bound will simply equal 1. 

\paragraph{Residual Imbalance.} The second component of the bias bound is the residual imbalance in an omitted confounder, and is a function of the $R^2$ parameter (Equation \ref{eqn:optim_bounds}-(b)). Similar to a point made in \cite{cinelli2020making}, there exists an asymmetry in the drivers of bias. More specifically, as the correlation between the imbalance term and the outcome increases towards 1 (i.e., the correlation bound approaches 1), the overall impact on the bias bound is bounded at 1.  In contrast, as the level of imbalance in the omitted confounder increases, the effect on the bias bounds is unbounded. In other words,  as $R^2 \to 1$, the corresponding bias bounds will increase towards infinity. 

\paragraph{Scaling Factor.} The last factor in the bias bound is a scaling factor (represented by Equation \eqref{eqn:optim_bounds}-(c)). The scaling factor comprises of the variance of the outcomes across the control units (i.e., $\var(Y_i \mid Z_i = 0)$) and the variance of the estimated weights (i.e., $\var(w_i \mid Z_i = 0)$). This represents the overall heterogeneity that is present in the data. More specifically, as the variance in the estimated weights increases, there is more imbalance between the treatment and control groups that the weights are accounting for. Similarly, as the variance in the outcomes increases, there is more potential for heterogeneity to be related to the selection into treatment, making it more difficult to recover the true estimated effect.\footnote{We note that in settings when $\var(Y_i \mid Z_i=0) = 0$, there is \textit{no} variation in the outcomes across the control group. As such, no amount of weighting will alter our estimate.} The scaling factor is a function of the observed data, and is not related to the omitted confounder. However, the scaling factor serves as an amplification of any bias that would arise from omitting a confounder.

The key takeaway from Theorem \ref{thm:optim_bounds} is that given a researcher-chosen $R^2$ value, the optimal bias bound in Theorem \ref{thm:optim_bounds} is directly estimable from the data. As such, for a fixed $R^2$ value, researchers can directly calculate the range of possible bias values.

\subsection{Constructing Confidence Intervals} \label{subsec:valid_ints}  
We now introduce a method to construct valid asymptotic confidence intervals for the variance-based sensitivity models. Our method builds on the work of \cite{zhao2019sensitivity} and uses a percentile bootstrap to simultaneously accounts for the bias due to omitting a confounder and the uncertainty associated with estimation. Our approach is distinct from those in the partial identification literature that require known asymptotic distributions of the boundaries of the partially identified region  \citep{imbens2004confidence, aronow2013interval}; similar to the discussion provided in \cite{zhao2019sensitivity}, it can difficult to characterize these distributions analytically in our sensitivity framework. Instead, the proposed bootstrap approach allows researchers to account for sampling uncertainty without explicitly characterizing the asymptotic distributions of the boundary estimates. 

To begin, for a fixed $R^2$, we can define $\hat \tau(\tilde w)$ as the weighted estimate, using weights $\tilde w \in \sigma(R^2)$. We can equivalently view $\hat \tau(\tilde w)$ as an adjusted weighted estimate for some $\tilde w \in \sigma(R^2)$: 
\begin{equation} 
\hat \tau(\tilde w) := \hat \tau_W - \text{Bias}(\hat \tau_W \mid \tilde w),
\label{eqn:adj_est} 
\end{equation} 
where $\text{Bias}(\hat \tau_W \mid \tilde w)$ is the bias of $\hat \tau_W$, assuming $\tilde w$ were the true weights (i.e., $\text{Bias}(\hat \tau_W \mid \tilde w) :=  \hat \tau_W - \hat \tau(\tilde w)$), assuming the true weights are equal to $\tilde w$. 

Applying the results from \cite{zhao2019sensitivity}, for every $\tilde w \in \sigma(R^2)$, we can construct a confidence interval for $\tau(\tilde w)$ using a percentile bootstrap: 
\begin{equation} 
[L(\tilde w), U(\tilde w)] = \left[ Q_{\alpha/2}\big(\hat \tau^{(b)}(\tilde w) \big), Q_{1-\alpha/2}\big(\hat \tau^{(b)}(\tilde w) \big) \right],
\label{eqn:def_bounds} 
\end{equation} 
where $\hat \tau^{(b)}(\tilde w)$ is the adjusted weighted estimator in bootstrap sample $b \in \{1, ..., B\}$, and $Q_\alpha(\cdot)$ denotes the $\alpha$-th percentile in the bootstrap distribution. By the following theorem, $[L(\tilde w), U(\tilde w)]$ will be an asymptotically valid (1-$\alpha$) confidence interval for $\tau(\tilde w)$:% with at least $(1-\alpha)$ coverage: 
\begin{theorem}[Validity of Percentile Bootstrap] \mbox{} \label{thm:bs} \\
Under mild regularity conditions (see Assumption \ref{assump:regularity_conds} in the Appendix), for every $\tilde w \in \sigma(R^2)$: 
$$\limsup_{n \to \infty} P(\tau(\tilde w) < L(\tilde w)) \leq \frac{\alpha}{2} \text{ and } \limsup_{n \to \infty} P(\tau(\tilde w) > U(\tilde w)) \leq \frac{\alpha}{2},$$
where $L(\tilde w)$ and $U(\tilde w)$ are defined as the $\alpha/2$ and $1-\alpha/2$-th quantiles of the bootstrapped estimates (i.e., Equation \eqref{eqn:def_bounds}). 
\end{theorem} 
Theorem \ref{thm:bs} states that for any set of weights $\tilde w$, the percentile bootstrap can be applied to estimate valid confidence intervals for an adjusted, weighted estimate $\tau (\tilde w)$.  However, as discussed in the previous section, for a given $R^2$ value, there exists many different sets of weights $\tilde w$ that can be defined from $\sigma(R^2)$. As such, we apply the union method to estimate a conservative $(1-\alpha)$\% confidence interval $\text{CI}(\alpha)$ for $\tau(\tilde w)$: 
\begin{equation} 
\text{CI}(\alpha) = \left[Q_{\alpha/2} \left( \inf_{\tilde w \in \sigma(R^2)} \hat \tau^{(b)}(\tilde w)\right), \ Q_{1- \alpha/2} \left( \sup_{\tilde w \in \sigma(R^2)} \hat \tau^{(b)}(\tilde w) \right) \right].
\label{eqn:conf_int} 
\end{equation} 
We can estimate $\text{CI}(\alpha)$ directly from the bootstrap samples by calculating the minimum and maximum adjusted weighted estimate for each bootstrap iteration, and then estimating the $\alpha/2$ and $1-\alpha/2$-th percentiles across the bootstrap distributions. The extrema of the adjusted weighted estimate follow directly from the results of Theorem \ref{thm:optim_bounds}:
$$\inf_{\tilde w \in \sigma(R^2)} \hat \tau(\tilde w) = \hat \tau_W - \max_{\tilde w \in \sigma(R^2)} \text{Bias}(\hat \tau_W \mid \tilde w) \ \ \ \ \ \sup_{\tilde w \in \sigma(R^2)} \hat \tau(\tilde w) = \hat \tau_W + \max_{\tilde w \in \sigma(R^2)} \text{Bias}(\hat \tau_W \mid \tilde w)$$

As such, to estimate valid confidence intervals, researchers can use a percentile bootstrap, estimating the bias bound and calculate an adjusted point estimate in each bootstrap sample. We summarize the steps in Figure \ref{fig:steps}. 

\begin{figure}[ht]
{
\singlespacing
\noindent\fbox{%
\vspace{2mm}
\parbox{0.98\textwidth}{%
{
\vspace{2mm}
\noindent \underline{\textbf{Valid Confidence Intervals for the Variance-Based Sensitivity Model}}

\begin{Step} 
\item Fix $R^2 \in [0,1)$ and generate $B$ bootstrap samples of the data. 
\item For each bootstrap sample $b = 1, ..., B$:  
\begin{enumerate} 
\item Estimate weights $\hat w_i^{(b)}$ and the point estimate $\hat \tau_W^{(b)}$. 
\item Calculate $\widehat{\var}_b(Y_i)$, $\widehat{\cor}_b(\hat w_i^{(b)}, Y_i)$, and $\widehat{\var}_b(\hat w_i^{(b)})$, where the subscript $b$ denotes the quantity calculated over the $b$-th bootstrap sample.
\item Use the optimal bias bounds (i.e., Equation \eqref{eqn:optim_bounds}) to calculate the range of potential point estimates: 
$$\hat \tau^{(b)}(\tilde w) \in \Bigg[\hat \tau^{(b)}_W - \max_{\tilde w \in \sigma(R^2)} \text{Bias}(\hat \tau^{(b)}_W \mid \tilde w), \ \  \hat \tau^{(b)}_W + \max_{\tilde w \in \sigma(R^2)} \text{Bias}(\hat \tau^{(b)}_W \mid \tilde w)\Bigg]$$
\end{enumerate} 
\item From the $B$ bootstrapped optimal bounds, estimate the $\alpha/2$ and $1-\alpha/2$-th percentiles of the minima and maxima values respectively to obtain valid confidence intervals (i.e., Equation \eqref{eqn:conf_int}). 
\end{Step} 
} 
} }\\
\vspace{2mm}
}
\caption{Summary of percentile bootstrap procedure for estimating valid confidence intervals.} \label{fig:steps}
\end{figure}

\subsection{Conducting the Sensitivity Analysis} \label{subsec:benchmarking} 
To conduct the sensitivity analysis, researchers estimate confidence intervals for increasing $R^2$ values until an estimated confidence interval just contains the null estimate; the corresponding $R^2$ is denoted as $R^2_*$. Because the $R^2$ is bounded on an interval $[0,1]$, researchers are restricted by the range of values that they can posit for the different $R^2$ values. However, it can nonetheless be difficult to reason about the plausibility of a given $R^2$ value and the strength or weakness of confounders on the $R^2$ scale. 

Previous papers have suggested the use of benchmarking to assess what may be plausible sensitivity parameters \citep{huang2022sensitivity, hartman2022survey, cinelli2020making, hong2021did, carnegie2016assessing, hsu2013calibrating}. To perform benchmarking, researchers sequentially omit different observed covariates and re-estimate the weights. They can then directly calculate the error that arises from omitting each covariate and directly estimate the sensitivity parameters (or bounds for the sensitivity parameters). The estimated sensitivity parameters from omitting each covariate are then interpreted as the sensitivity parameters for omitted confounders with equivalent confounding strength to an observed covariate. If researchers have a strong substantive understanding of covariates that explain a lot of the variation in the treatment assignment mechanism or outcome, then benchmarking can be a useful tool for understanding the strength of hypothetical omitted variables.

We propose a formal benchmarking procedure for the variance-based sensitivity model. To begin, let there be $p$ total observed covariates (i.e., $\bX \in \R^{n \times p}$). Then for the $j$-th covariate, where $j \in \{1, ..., p\}$, we define the benchmarked weights $w^{-(j)}$ as the estimated weights, containing all covariates, except for the $j$-th covariate. Using $w^{-(j)}$, we can estimate the benchmarked $R^2$ value for an omitted confounder that is equivalently imbalanced as the $j$-th covariate: 
\begin{equation} 
\hat R^2_{(j)} = \frac{\hat R^2_{-(j)}}{1+\hat R^2_{-(j)}}, \ \ \ \ \ \text{where } \hat R^2_{-(j)} := 1 - \frac{\var(w_i^{-(j)})}{\var(w_i)}.
\label{eqn:r2_benchmark} 
\end{equation} 
$\hat R^2_{(j)}$ represents the $R^2$ value of an omitted variable that has the same amount of residual imbalance as the $j$-th covariate.\footnote{The reason we cannot directly use $\hat R^2_{-(j)}$ as an estimate for the benchmarked $R^2$ value comes from the fact that we must adjust for changes in the baseline variation of the weights between $w$ and $w^*$. We refer readers to \cite{cinelli2020making} and \cite{huang2022sensitivity} for more discussion on this point.} More specifically, $R^2_{(j)}$ corresponds to an omitted variable with the same amount of imbalance, after controlling for $\bX$, as the $j$-th covariate, after controlling for $\bX^{-(j)}$. 

When interpreting the benchmarking results, it is important to consider that the magnitude of the benchmarked $R^2$ values is determined by the \textit{residual} imbalance. More concretely, we consider the variables \textit{income} and \textit{education} in the running example. We expect that both income and education will be predictive of individuals’ propensity for fish consumption. However, omitting just income may not result in a very large $R^2$ value, because by balancing education, we have implicitly controlled for some of the imbalance in income. The benchmarked $R^2$ parameter thus represents the setting in which researchers have omitted a variable that, when controlling for all the other observed variables, has the same amount of residual imbalance as income after controlling for education. In cases when researchers wish to consider omitting a variable similar to a set of collinear variables, they can omit subsets of variables and perform the same benchmarking exercise.

Formal benchmarking can also be used to assess the plausibility of the event $R^2  \geq R^2_*$. More specifically, we can directly compare the benchmarked $\hat R^2_{(j)}$ values for $j \in \{1,...,p\}$ with the estimated $R^2_*$ to see how much more or less imbalanced an omitted confounder must be, relative to an observed covariate, in order to result in an $R^2$ value equal to $R^2_*$. We refer to this as the \textit{minimum relative imbalance} (MRI): 
$$\text{MRI}(j) = \frac{R^2_*}{\hat R^2_{(j)}}.$$
If the MRI is small (i.e., $\text{MRI}(j) < 1$), the omitted confounder need not be very imbalanced, relative to the $j$-th covariate, in order to make a null result plausible. In contrast, if the MRI is large (i.e., $\text{MRI}(j) > 1$), then the omitted confounder must be more imbalanced than the $j$-th observed covariate to make a null result plausible. 

Formal benchmarking offers an opportunity for researchers to incorporate their substantive understanding into the sensitivity analysis and provides much-needed interpretability for the sensitivity framework. In particular, when researchers have strong priors about which underlying observed variables control the treatment assignment mechanism, formal benchmarking is very useful for reasoning about the plausibility of an omitted confounder strong enough to explain observed results in the absence of a true effect.

\subsection{Illustration on NHANES} \label{subsec:vbm_nhanes}
In our running example, we begin by varying the $R^2$ parameter across the range $[0, 1)$, and estimate the corresponding the 95\% confidence intervals. We estimate $R^2_* = 0.57$, such that if $R^2 \geq 0.57$, the intervals contain the null estimate. This implies that if the omitted confounder explains 57\% or more of the variation in the true weights, our estimated effect of fish consumption on blood mercury levels is no longer significantly different from the expected distribution under the null.

To assess the plausibility of an omitted confounder resulting in an $R^2$ value of 0.57, we perform formal benchmarking and estimate  benchmarked $\hat R^2$ values for each covariate. Omitting a confounder like race, education, or income results in the largest $R^2$ values. More specifically, omitting a confounder with equivalent confounding strength to race results in an $R^2$ of 0.19, while omitting a confounder with equivalent confounding strength to education or income results in an $R^2$ of 0.17 and 0.14, respectively. From these results, we see that an omitted confounder would have to explain around 3 times the variation in true weights as the strongest observed covariate, race, in order for the $R^2$ value to equal the cutoff value. We argue that while mathematically possible, the plausibility of a confounder resulting in the threshold $R^2_* = 0.57$ value is low.

\begin{figure}[!ht]
\centering
\includegraphics[width=0.9\textwidth]{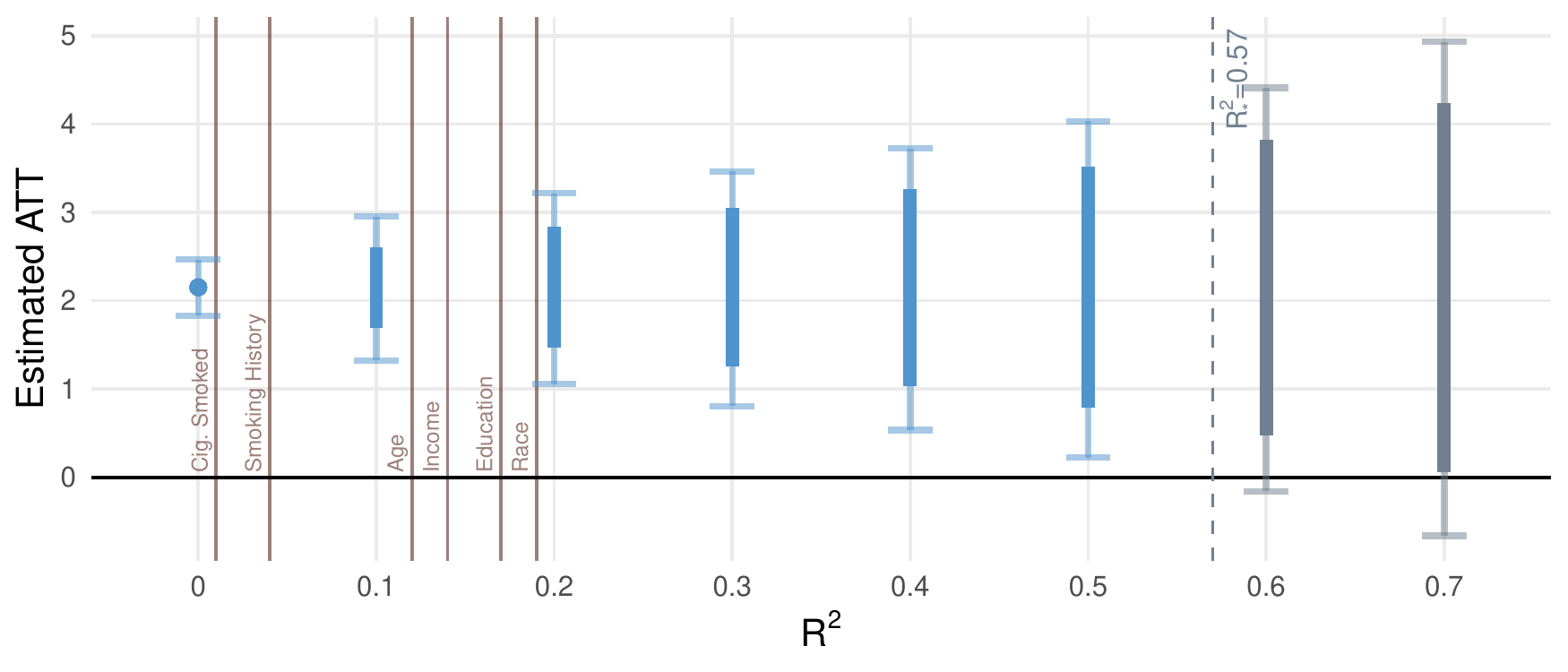}
\caption{Results from the sensitivity analysis under the variance-based sensitivity model. We vary the $R^2$ measure across the $x$-axis and plot the range of estimated ATT values on the $y$-axis. The solid bar denotes the point estimate bounds for a specified $R^2$ value, estimated as the point estimate plus and minus the optimal bias bounds (Theorem \ref{thm:optim_bounds}). The lighter intervals represent the 95\% confidence intervals. We also plot the benchmarking results for the observed covariates, where the lines represent the corresponding benchmarked $R^2$ values.} 
\end{figure}

\subsection{Extension: Bounding a Confounder's Relationship with the Outcome} \label{subsec:ext_corr}
Previous literature has highlighted two characteristics of the imbalance term in Equation \eqref{eqn:w_decomp} that affect the bias from omitting a variable: (1) the overall magnitude of the the imbalance term, and (2) the relationship between the imbalance term to the outcomes (e.g., \cite{huang2022sensitivity, hong2021did, cinelli2020making, shen2011sensitivity}). Like the marginal sensitivity model, the variance-based sensitivity model constrain the overall magnitude of the imbalance term, and implicitly assume that the imbalance is maximally correlated with the outcome. In settings when researchers wish to account for this additional characteristic of the imbalance term, the variance-based sensitivity model can be easily extended to allow researchers to bound the relationship between the imbalance and the outcome. In particular, unlike the marginal sensitivity model, in which researchers must solve a linear programming problem to identify the extrema, there exists a closed-form solution for the optimal bias bounds under the variance-based sensitivity model. As such, researchers can choose to evaluate the optimal bias bounds and associated confidence intervals using less conservative values of the correlation bound.

While amplification approaches for examining the relationship between the outcomes and the confounder for a fixed level of imbalance exist for alternative sensitivity models, many of these methods require introducing additional complexities. (For example, \cite{rosenbaum2009amplification} requires invoking parametric assumptions on the outcomes.) In contrast, the variance-based sensitivity model allow researchers to easily incorporate additional information about the confounder to directly bound the relationship between the outcome and the imbalance in an omitted confounder. We provide recommendations for alternative bounds that researchers can use in Appendix \ref{sec:relax_rho}, as well as benchmarking procedure that allows researchers to use observed covariate data to estimate plausible correlation bounds. 

\section{Relationship to the Marginal Sensitivity Model} \label{sec:msm_compare} 
We now examine the relationship between the variance-based sensitivity model and the marginal sensitivity model. We first show that both sets of sensitivity models can be written as norm-constrained optimization problems. The variance-based sensitivity model implicitly constrains a weighted $L_2$ norm, while the marginal sensitivity model constrains an $L_\infty$ norm. We demonstrate that by moving away from a worst-case characterization of the error from an omitted variable, the variance-based sensitivity model can obtain narrower, more informative bounds. We illustrate the potential for narrower bounds using benchmarked results from the running example. 

\subsection{Sensitivity Models as an Optimization Problem} \label{subsec:norm} 
Re-formulating the variance-based sensitivity model as an optimization problem under a bounded norm enables comparison with the marginal sensitivity model, which can be written as bias maximization problems under a constrained $L_\infty$ norm. We argue that constraining the weighted $L_2$ norm can produce less conservative estimated bounds.

To begin, we show that the variance-based sensitivity model is a bias maximization problem, given a fixed constraint on a weighted $L_2$ norm. 
\begin{theorem}[Weighted $L_2$ Norm Constraint] \label{thm:l2_norm} \mbox{}\\
Define the individual-level error in the weights as $\lambda_i := w_i^*/w_i$. Define the $L_{2,w}$ norm as follows: 
$$||\lambda||^2_{2,w} := \begin{cases} \displaystyle 
\frac{1}{n} \sum_{i=1}^n \lambda_i^2 \cdot \nu(w_i) & \text{if } \var(w_i) > 0, \\
\infty & \text{else}
\end{cases},$$
where $\nu(w_i)$ is a function of the estimated weights. Then, the variance-based sensitivity model can equivalently be written as a norm-constrained optimization problem:
$$\max_{\tilde w \in \sigma(R^2)} \text{Bias}(\hat \tau_W \mid \tilde w) \iff \begin{cases} \displaystyle \max_{\tilde w} \text{Bias}(\hat \tau_W \mid \tilde w) \\ 
\displaystyle \text{s.t. } ||\lambda||_{2,w} \leq \sqrt{\frac{k}{1-R^2}},
\end{cases}$$
where $k := 1-R^2/\E(w_i^2)$. See Appendix \ref{app:proofs} for proof and details. 
\end{theorem} 

Theorem \ref{thm:l2_norm} is especially in concert with the following result from \cite{zhao2019sensitivity} showing that the marginal sensitivity model is constrained $L_\infty$ problems: 
$$\max_{\tilde w \in \varepsilon(\Lambda)} \text{Bias}(\hat \tau_W \mid \tilde w) \iff \begin{cases} 
 \displaystyle \max_{\tilde w} \ \ \text{Bias}(\hat \tau_W \mid \tilde w) \\ 
\displaystyle \text{s.t. } \Lambda^{-1} \leq  ||\lambda||_{\infty} \leq \Lambda.
\end{cases}$$
These constrained-norm representations provide insight into the benefits expected from the variance-based sensitivity model. Because the marginal sensitivity model  optimize over the set of weights defined by a worst-case error, the estimated bounds on the bias always correspond to cases in which \textit{all} units are exposed to this worst-case error. However, in settings when one or two subjects are subject to much larger levels of confounding than others, this can result in an overly pessimistic view of the potential bias \citep{fogarty2019extended, zhao2019sensitivity}. In contrast, the variance-based sensitivity model is optimizing over a set of weights defined by average weighted error, and thus allow a small number of weights to be exposed to large amounts of error, even at moderate levels of overall confounding.

\subsection{Comparison of Estimated Bounds} \label{subsec:oracle}
While the constrained-norm representation provides intuition for why the variance-based sensitivity model may obtain narrower bounds than the marginal sensitivity model, in practice it is difficult to directly compare the bounds estimated under the two families of models.  This is because the two approaches are using two different parameters and are fundamentally characterizing the error from omitting a confounder in a different manner. In the following subsection, we consider a setting in which researchers can estimate bounds using the true sensitivity parameter and compare the size of the associated confidence intervals. While in practice, researchers do not have access to the true sensitivity parameters, this approach provides intuition for the relative performances of the two sensitivity models. 

First, we formalize a condition under which the variance-based sensitivity model will result in narrower bounds than the marginal sensitivity model. In general, we expect the variance-based sensitivity model to result in narrower bounds if the worst-case error ( $\Lambda$) is much larger than the true average weighted error (proxied by $R^2$). If the difference in the worst-case error and average weighted error is not very large, then there will not be much improvement in the estimated bounds from using the variance-based sensitivity model. Theorem \ref{thm:bound_cond} provides a maximum threshold for the size of $R^2$  relative to the $\Lambda$ value sufficient for strictly narrower bounds under the variance-based sensitivity model. 
%\clearpage
\begin{theorem}[Narrower Bounds under the Variance-Based Sensitivity Model] \label{thm:bound_cond}
Let $\psi(\Lambda)$ represent the difference in the estimated point estimate bounds under the marginal sensitivity model $\varepsilon(\Lambda)$ for a given $\Lambda \geq 1$: 
$$\psi(\Lambda):= \max_{\tilde w \in \varepsilon(\Lambda)} \frac{\sum_{i:Z_i = 0} Y_i Z_i \tilde w_i}{\sum_{i:Z_i=0} Z_i \tilde w_i}- \min_{\tilde w \in \varepsilon(\Lambda)}\frac{\sum_{i:Z_i = 0} Y_i Z_i \tilde w_i}{\sum_{i:Z_i=0} Z_i \tilde w_i}.$$ 
Then if the true $R^2$ parameter is lower than the following threshold,
\begin{equation} 
R^2 \leq \frac{\psi(\Lambda)^2}{4 \underbrace{(1-\cor(w_i, Y_i \mid Z_i = 0)^2)}_{\text{Correlation Bound}} \cdot \underbrace{\var(w_i\mid Z_i = 0) \var(Y_i\mid Z_i = 0)}_{\text{Scaling Factor}} + \psi(\Lambda)^2},
\label{eqn:cond} 
\end{equation} 
the bounds under the variance-based sensitivity model will be narrower than the bounds for the marginal sensitivity model.
\end{theorem} 

 Besides the worst-case error $\Lambda$, the $R^2$ threshold is determined by the correlation between the estimated weights and the outcome, $\cor(w_i, Y_i\mid Z_i = 0)$, and the scaling factor, $\var(w_i\mid Z_i = 0) \cdot \var(Y_i\mid Z_i = 0)$. These components affect the estimated bounds under both sets of sensitivity models. Both $\cor(w_i, Y_i\mid Z_i = 0)$ and the scaling factor are direct inputs into the optimal bias bounds under the variance-based sensitivity model. In addition, increases in these quantities either lead to larger outcome values, or more extreme weights; because the optimal bounds under the marginal sensitivity model are estimated by scaling the weights and outcomes by $\Lambda$ (or $\Lambda^{-1}$), $\psi(\Lambda)$ will also increase.

Theorem \ref{thm:bound_cond} does not guarantee that the variance-based sensitivity model will always result in narrower bounds than the marginal sensitivity model, but we consider two specific scenarios that highlight the practical advantages from using variance-based sensitivity model. First, we consider an asymptotic setting. We show that in many cases, the worst-case error $\Lambda$ will diverge to infinity, regardless of the omitted variable’s confounding strength. In contrast, the $R^2$ parameter is a direct function of the confounding strength and retains a more stable interpretation across different data scales.  

Second, we consider a finite-sample setting, in which the outcomes and probability of treatment are highly correlated (which we refer to as \textit{limited outcome overlap}). In this setting, the marginal sensitivity model may produce narrower intervals, but these intervals can be misleadingly narrow and fail to provide nominal coverage. In contrast, while the intervals under the variance-based sensitivity model will be wider in this setting, the intervals adequately account for the limited outcome overlap and continue to provide nominal coverage. 

\paragraph{Remark.} Several recent extensions of the marginal sensitivity model allow researchers to mitigate some the conservative nature of the method by adding in additional constraints beyond bounding the worst-case error  \citep{kallus2018confounding, dorn2021sharp, dorn2021doubly}. However, these methods usually require adding  additional sensitivity parameters or performing some form of outcome modeling. The variance-based sensitivity model offers stable and informative bounds via a one-parameter sensitivity analysis without additional assumptions, constraints, or complexities.

\subsubsection{Infinite Worst-Case Error in Asymptotic Settings} 
We begin by considering the asymptotic setting. We show that when the omitted confounder results in an error that can be arbitrarily small or large and the outcomes $Y_i$ are unbounded, the asymptotic confidence intervals for the variance-based sensitivity model will necessarily be narrower than the confidence intervals estimated under the marginal sensitivity model. Consider the following instructive example. 
\begin{example}[Behavior of $\Lambda$ for a Logit Model] \label{ex:infty_lambda} \mbox{}\\
Assume researchers use a logit model to estimate the weights using $\bX_i$, but omit a confounder $U_i$. The estimated and ideal weights take on the following forms: 
\begin{equation*} 
\hat w_i = \exp(\hat \gamma^\top \bX_i) \ \ \ \ \ \ \ \ \ \ \hat w_i^* = \exp(\hat \gamma^{*\top} \bX_i + \hat \beta U_i)
\end{equation*} 
Then let $\hat \Lambda$ be the maximum error across our observed sample (i.e., $\hat \Lambda := \max_{1\leq i\leq n} \{\hat w_i^*/\hat w_i, \hat w_i/\hat w_i^* \}$). Assume $\begin{bmatrix} \bX_i,U_i\end{bmatrix} \iid MVN(0, I)$. Then $\E(\hat \Lambda) \to \infty$ as $n \to \infty$: 
$$\lim_{n \to \infty} \frac{\E(\hat \Lambda)}{\exp(\sqrt{2 \nu^2 \log(n)})} \geq 1,$$
where $\nu^2 = (\gamma^* - \gamma)^\top (\gamma^* - \gamma) + \beta^2$ (where $\gamma^*$, $\gamma$, and $\beta$ represent the population counterparts to the estimated coefficients), and the results follow immediately from \cite{wainwright2019high}. 
\end{example} 

Example \ref{ex:infty_lambda} highlights that the worst-case error will increase towards infinity as the sample size grows, \textit{regardless} of the confounding strength of the omitted variable (represented by $\hat \beta$). The omitted confounder could explain little of the treatment assignment process, but the worst-case bound on the error would still be infinitely large. This means that researchers would have to specify an infinitely large $\Lambda$ value for the marginal sensitivity model to be valid. In other words, while we can find a $\hat \Lambda$ in an observed sample that provides an upper bound on the difference between the realized and the ideal weights, the marginal sensitivity model will not hold for \textit{any} value of $\Lambda$ in the population \citep{jin2022sensitivity}. Intuitively, this occurs because $\hat{\Lambda}$ is a function of the largest $U_i$ value in the sample: as the sample size increases, the probability of observing an extreme $U_i$ value increases. 

A secondary issue arises from the decoupling of the magnitude of the sensitivity parameter and the underlying confounding strength of the omitted variable. In particular, reasoning about whether or not $\Lambda$ values is large or small is can be challenging. Even with the aid of benchmarking, researchers can at best estimate the worst-case error that arises from omitting different covariates, but reasoning about whether or not it is plausible for such an error to arise from an omitted variable amounts to reasoning about whether or not it is plausible for potential outliers to occur. Furthermore, as the sample size increases, researchers must take into account the potential for more outliers to occur. 

In contrast, we can calculate the $R^2$ value for the variance-based sensitivity model under the same setting as Example \ref{ex:infty_lambda}. It is a function of $\hat \gamma$, $\hat \gamma^*$, and $\hat \beta$ that does not depend on the sample size. 
\begin{example}[Behavior of $R^2$ for a Logit Model] \label{ex:r2_logit} \mbox{}\\
Consider the same setting as Example \ref{ex:infty_lambda}. Then, the $R^2$ value can be written as follows: 
$$R^2 = 1 - \frac{\exp(\hat \gamma^\top \hat \gamma) - 1}{\exp(\hat \gamma^{*\top} \hat \gamma^* + \hat \beta^2) - 1} \cdot \frac{\exp(\hat \gamma^\top \hat \gamma)}{\exp(\hat \gamma^{*\top} \hat \gamma^* + \hat \beta^2)}.$$
\end{example} 

Example \ref{ex:infty_lambda} gives one setting in which $\Lambda$ will be infinitely large, regardless of the confounding strength of the omitted variable. More generally, the following corollary to Theorem \ref{thm:bound_cond} shows that 
under any setting when the error from omitting a confounder can take on values that are arbitrarily small or large, the  variance-based sensitivity model necessarily produces narrower bounds in sufficiently large samples. 

\begin{corollary}[Narrower Bounds under the Variance-Based Sensitivity Model] \label{cor:asy_vsm} 
Consider the set of confounders, in which for all $\delta > 0$, $P(w^*_i / w_i < \delta) > 0$, or $P(w^*_i / w_i > \delta) > 0$. Then, if the outcomes are unbounded, $\psi(\Lambda)$ will diverge in probability to infinity, and the threshold from Theorem \ref{thm:bound_cond} will converge in probability to 1:
$$\frac{\psi(\Lambda)^2}{4 (1-\cor(w_i, Y_i)^2) \cdot \var(w_i) \var(Y_i) + \psi(\Lambda)^2} \cip 1$$
\end{corollary}
Because $R^2 < 1$ by definition, for sufficiently large $n$, the variance-based sensitivity model will produce narrower bounds.

Corollary \ref{cor:asy_vsm} highlights that in certain asymptotic settings, if the outcomes are unbounded, the marginal sensitivity models will result in infinitely large bias bounds. This will occur, regardless of whether the omitted confounder is strong or weak. In contrast, because the variance-based sensitivity model is not using a worst-case characterization of error, the resulting bias bounds will be less susceptible to extreme values and be narrower by construction. 

\subsubsection{Limited Overlap in Finite-Samples} 
We now consider a finite-sample setting. We show that paradoxically, in certain finite-samples, the marginal sensitivity model can result in narrower bounds than the variance-based sensitivity model even when $\Lambda$ is very large. However, the narrower bounds come with a risk of substantial undercoverage, especially when the sample size is small or there is limited outcome overlap. In these settings, the variance-based sensitivity model will tend to return wider intervals, but maintain nominal coverage. 

The key to this phenomenon is a property of the marginal sensitivity model, referred to as \textit{sample boundedness}. Sample boundedness implies that even at infinitely large $\Lambda$ values, the worst-case bounds under the marginal sensitivity model approach, but cannot exceed, the range of the observed control outcomes (i.e., $\lim_{\Lambda \to \infty} \psi(\Lambda) \leq \max_{i:Z_i = 0} Y_i - \min_{i:Z_i = 0} Y_i$).

In contrast, the variance-based sensitivity model is not inherently sample bounded. In settings with relatively large amounts of confounding, the marginal sensitivity model will have narrower intervals than the variance-based sensitivity model, since as $R^2$ increases towards 1, the estimated bounds under the variance-based sensitivity model will be adequately wide.  However, sample boundedness may prohibit the construction of valid confidence intervals unless in the absence of a key implicit assumption on the distribution of the unobserved potential outcomes. Consider the following toy example: 

\begin{example}[Misleading Optimism from Sample Boundedness] \mbox{}\\
Consider the following population of 4 units, with the following potential outcomes, treatment assignment, and the estimated probability of treatments for each unit: 
%\clearpage
\begin{table}[!ht]
\centering 
\begin{tabular}{l|cccc}
$i$ & $Y_i(0)$ & $Y_i(1)$ & $\hat P(Z_i = 1)$ & $Z_i$\\ \hline
1  & -10 & \textcolor{lightgray}{-10} & 0.1 & 0\\ 
2 & 5 & \textcolor{lightgray}{5} & 0.2  & 0 \\ 
3 & \textcolor{lightgray}{10} & 10 & 0.9 & 1\\
4 & \textcolor{lightgray}{20} & 20 & 0.95 & 1
\end{tabular} 
\end{table} 

\noindent The true ATT is zero, but the estimated ATT is equal to 14.6, so substantial confounding is present.  However, since the sample bounds for the ATT are the interval $[10,25]$, no value of $\Lambda$ can produce an estimated interval (under the marginal sensitivity model) containing zero, erroneously suggesting the presence of a true effect highly robust to substantial confounding.

\end{example} 
While this example is somewhat contrived, it highlights the problems with sample boundedness if the potential outcome ranges in the two groups have limited overlap, which may occur when potential outcomes are strongly correlated with the probability of treatment.  For a formal characterization of this outcome overlap condition, see Appendix \ref{app:overlap}. 

When there exists limited outcome overlap, estimated intervals from the marginal sensitivity model may be misleadingly optimistic, especially for dramatic levels of potential confounding, but intervals constructed under the variance-based sensitivity model, which are not sample bounded, are not affected. 

\begin{figure}[ht] 
\centering 
\includegraphics[width=0.95\textwidth]{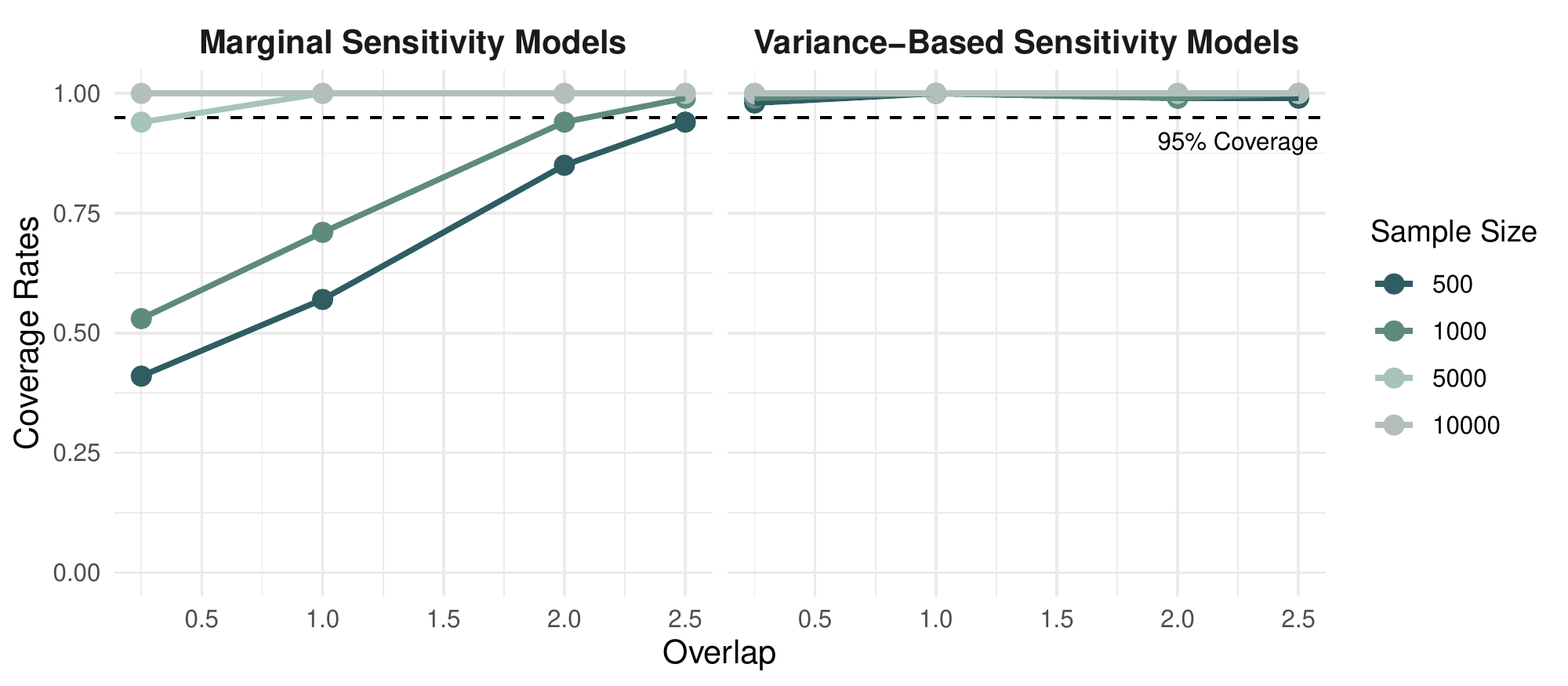}
\caption{Coverage rates for the marginal sensitivity model and the variance-based sensitivity model, assuming an oracle bias setting when researchers have full knowledge of the true underlying sensitivity parameter. The $x$-axis represents the amount of outcome overlap between the treatment and control groups (i.e., as $\sigma^2_v$, there is more outcome overlap).}
\label{fig:coverage}
\end{figure}

Figure \ref{fig:coverage} illustrates the behavior and coverage rates of both sets of sensitivity models under varying amounts of outcome overlap and sample sizes in an empirical example, described in greater detail in Appendix \ref{app:overlap}.

\paragraph{Remark.} We note that sample boundedness is not necessarily a negative feature in the context of \textit{estimation}. The bias-variance tradeoff of using a stabilized weighted estimator has been extensively studied (e.g., \cite{robins2007comment}). However, in the context of a sensitivity analysis, in which we are explicitly interested in examining the potential bias that can arise under varying levels of confounding, requiring sample boundedness can lead to misleading conclusions, and potential issues with outcome overlap should be considered carefully when interpreting results.

\subsection{Illustration on NHANES}
\begin{figure}[!ht]
\includegraphics[width=\textwidth]{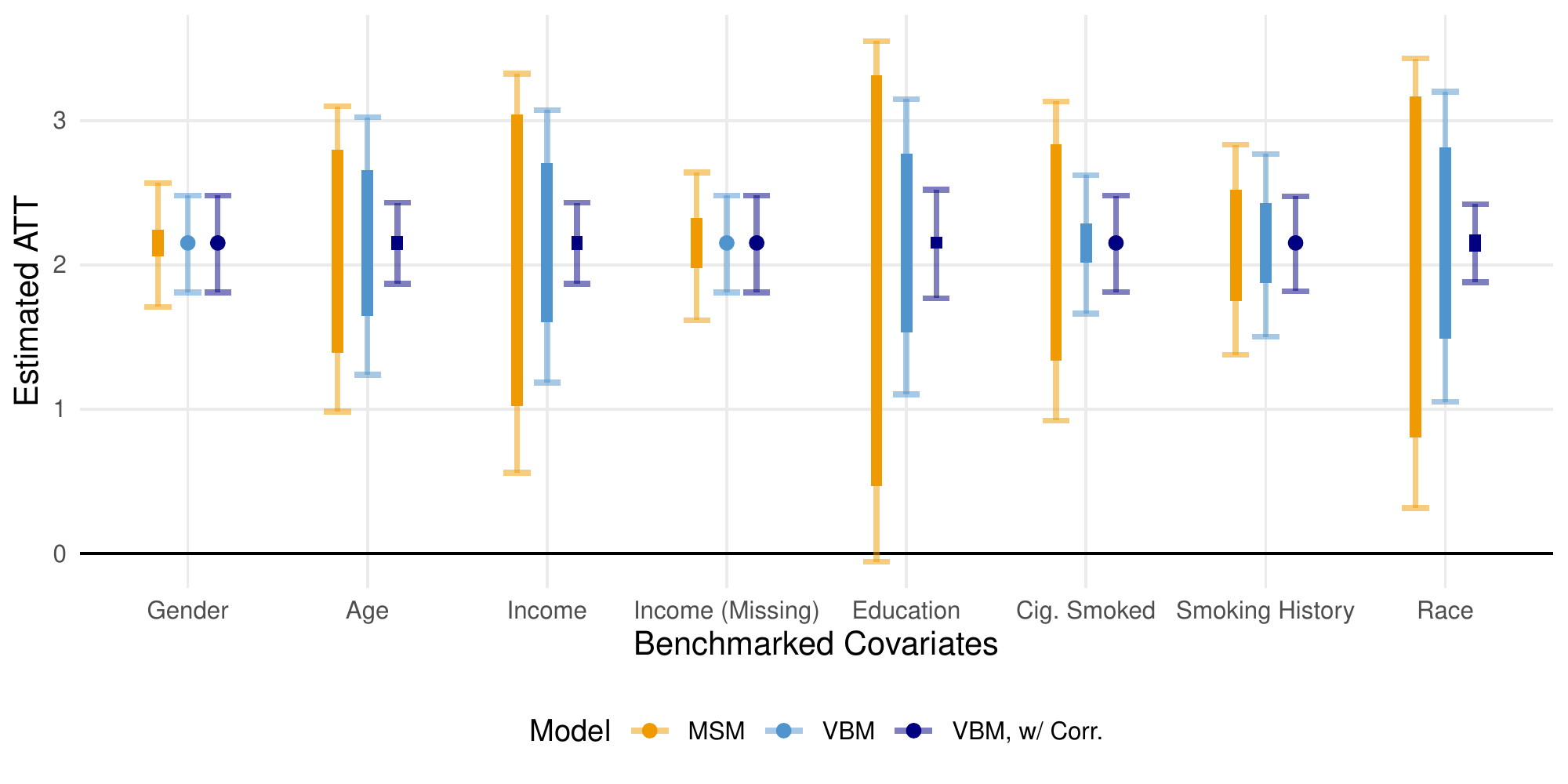}
\caption{Estimated intervals for both the marginal sensitivity model (in yellow) and the variance-based sensitivity model (in light blue), and the variance-based sensitivity model, using a less conservative correlation bound (in dark blue). The darker intervals represent the point estimate bounds, while the lighter intervals represent the 95\% confidence intervals. The intervals are estimated using the benchmarked $\Lambda$ and $R^2$ values for each covariate, and are interpreted as the resulting intervals for an omitted confounder with equivalent confounding strength to the observed covariate.} 
\label{fig:benchmark} 
\end{figure}

We now conduct formal benchmarking for the variance-based models and the marginal sensitivity model in our running example. We then estimate the corresponding bounds and intervals under both approaches. See Figure \ref{fig:benchmark} for a visualization. We see that for each of the covariates, omitting a confounder like any of the observed covariates would result in wider bounds under the marginal sensitivity model than the variance-based models. 

Notably, under the marginal sensitivity model, omitting a confounder like \textit{education} would be sufficient to explain the entire observed effect under the null hypothesis of no effect. In contrast,  under the variance-based sensitivity model, omitting a confounder with equivalent confounding strength to any of the observed covariates would not be sufficient to explain the observed data under the null. While the average error from omitting a variable like \textit{education} is relatively low, the maximum error that occurs is relatively large. The marginal sensitivity model, which assumes such a maximal error could occur in the unobserved confounder for most or all data points, thus produces much wider intervals than the variance-based model, which is much less responsive to individual outliers.

We also estimate intervals (and bounds) under the variance-based sensitivity model using a relaxed correlation bound. In particular, we choose the correlation bound by benchmarking an optional correlation parameter, giving the correlation between the outcome and the imbalance in an omitted confounder, to the observed correlation between the outcome and each observed covariate. (See Appendix \ref{sec:relax_rho} for more details.) By accounting for the relationship between the confounder and the outcome, we are able to obtain much narrower intervals. In particular, we see that even in cases where a potential omitted confounder is highly imbalanced (e.g., omitting a confounder like \textit{age} results in an $R^2$ value of 0.12, and $\Lambda$ value of 2.1), the overall bias that occurs from omitting it may be relatively low if the imbalance is largely unrelated to the outcome. By considering this additional dimension of the bias--which can be easily done using the variance-based sensitivity model-- researchers are able to better characterize the types of confounders that may lead to large amounts of bias and obtain a more holistic understanding of the sensitivity in their estimated effects. 

\section{Conclusion}

We have introduced a novel sensitivity model, the variance-based sensitivity model, which characterizes the error from omitting a confounder by using the distributional differences between the estimated weights and true weights. We show that the variance-based sensitivity model can be parameterized using an $R^2$ measure that represents the degree of residual imbalance in an omitted confounder, and provide methods for benchmarking the $R^2$ value of an omitted confounder against residual imbalances for observed covariates. We also derive a closed-form solution for the maximum possible bias and introduce a method for estimation of asymptotically valid confidence intervals under the sensitivity model.

The variance-based sensitivity model has several notable advantages over the existing marginal sensitivity model. First, by characterizing bias in terms of distributional differences instead of a worst-case error bound, variance-based sensitivity model can estimate narrower, less conservative bounds. Second, we show empirically that the variance-based sensitivity model obtains nominal coverage even in finite sample settings where the standard marginal sensitivity model dramatically undercovers due to issues with outcome overlap. Finally, because the variance based sensitivity model admits a closed-form solution for the optimal bias, we can introduce a natural two-parameter extension that uses constraints on the relationship between the omitted confounder and the outcome to produce narrower bounds. 

We suggest several directions for future work. First,  we demonstrated that variance-based sensitivity model, like the marginal sensitivity model, can be written as a norm-constrained optimization problem. Exploring other possible norms under which to constrain unobserved confounding could lead to a broad unified sensitivity framework, helping contextualize a wider variety of different sensitivity methods with their own strengths and weaknesses.

Second,  it is natural to ask what factors under a researcher's control at the design stage may influence the degree of robustness to unmeasured bias exhibited under the variance-based sensitivity analysis.  While the closed form for the optimal bias bound already provides insights in this direction, developing a metric akin to design sensitivity for matched studies \citep{rosenbaum2004design, rosenbaum2010designsensitivity} would provide valuable further guidance about how to design weighting estimators for maximum robustness. 

Finally, while we focused on a choice between bounding a weighted average error and bounding a worst-case error, future work could incorporate both constraints in the same study. We anticipate that this would result in further narrowing of sensitivity bounds. 

%\clearpage
%BIBLIOGRAPHY: 
\bibliographystyle{chicago} 
\bibliography{bibliography}
\clearpage 
 
\appendix
\setcounter{page}{1}
\begin{center}
    \singlespacing
    \Large
    \textbf{Supplementary Materials:} \\Variance-based Sensitivity Analysis for Weighting Estimators Result in More Informative Bounds
\end{center}
\singlespacing

\section{Additional Discussion} 
\subsection{Missingness} \label{app:missingness}  
In the man manuscript, the estimand of interest is the average treatment effect, across the treated. However, we note that the sensitivity framework introduced can be applied to more general settings, in which we consider missingness conditionally at random: 
$$Y_i \ \indep \ A_i \mid \mathcal{X}$$
This provides a very flexible framework to consider many settings of interest. Table \ref{tbl:missingness} summarizes several settings of interest, along with the associated conditional ignorability assumption to be relaxed by sensitivity analysis.
\begin{table}[!ht]
\begin{center} 
\begin{tabular}{lcc}
\toprule
Setting & Missingness Indicator  & Ignorability Statement\\ \midrule
Survey Response & $R_i$ (Response) & $Y_i \ \indep \ R_i \mid \mathcal{X}$\\ 
Internal Validity & $Z_i$ (Treatment Assignment) & $Y_i(1), Y_i(0) \ \indep \ Z_i \mid \mathcal{X}$\\ 
External Validity & $S_i$ (Inclusion in Experimental Sample) & $Y_i(1)-Y_i(0) \ \indep \ S_i \mid \mathcal{X}$ \\ \bottomrule
\end{tabular}
\caption{Summary of different common missingness settings.}
\label{tbl:missingness} 
\end{center} 
\end{table} 

\subsection{Relationship with Extensions for Sharper Bounds}
Recently, several papers have demonstrated that the worst-case bounds derived under the marginal sensitivity model result in $w^*$ that fail to recover the causal estimand. Thus, these worst-case bounds tend to be unnecessarily conservative, and may not necessarily be sharp. Both \cite{dorn2021sharp} and \cite{nie2021covariate} introduce additional optimization constraints for slightly tighter (and sometimes sharp) bounds. However, these additional optimization constraints come at a cost. In particular, the approach in \cite{dorn2021sharp} require imposing parametric assumptions of the conditional quantiles of the outcomes; furthermore, the method does not accommodate discrete outcome variables. \cite{nie2021covariate} include additional balancing constraints when solving for the bounds; however, \cite{dorn2021sharp} show that doing so can result in unstable performance in finite-sample settings. For this paper, we restrict our discussion to the marginal sensitivity model, but similar comparisons could be made to extensions of these models. Because these methods are all extensions of the marginal sensitivity model, the shortcomings and drawbacks that are discussed about the marginal sensitivity model similarly apply to these approaches as well.

\subsection{Parametric Assumption of Conditional Ignorability}\label{app:phiX} 
In practice, when researchers estimate weights, they are implicitly assuming a parametric version of Assumption \ref{assump:cond_ignor}. Following \cite{hartman2021kpop}, we formalize the parametric version of Assumption \ref{assump:cond_ignor}: 
\begin{assumption}[Linear ignorability in $\phi(\bX)$]\label{assump:linearign} \mbox{}\\
Let $\phi(\cdot)$ be a feature mapping of $\bX_i$. Then, write the outcome $Y_i$ as follows: 
$$Y_i =\phi(\bX_i)^{\top}\beta + \delta_i$$
Similarly, write $P(Z_i = 1 \mid \bX_i)$ as follows: 
\begin{align*}
Pr(Z_i=1|\bX_i) = g(\phi(\bX_i)^{\top}\theta + \eta_i),
\end{align*} 
where $g(\cdot):\mathcal{R}\mapsto [0,1]$. Then, linear ignorability holds when $\delta_i \indep \eta_i$.
\end{assumption}
Linear ignorability in $\phi(\bX_i)$ implies that the part of the outcome that is orthogonal to $\phi(\bX_i)$ is independent to the part of the treatment assignment process that is orthogonal to $\phi(\bX_i)$. 

The distinction between the non-parametric version  of conditional ignorability (i.e., Assumption \ref{assump:cond_ignor}) and the parametric version (i.e., Assumption \ref{assump:linearign}) arises from the types of violations that matter for omitted variable bias. Under the non-parametric version of conditional ignorability, only variables that are fully unobserved (or omitted) will result in bias. However, under Assumption \ref{assump:linearign}, in addition to including all of the correct variables, the choice of feature mapping also matters. For example, if researchers only include first-order moments in their weights estimation, then $\phi(\bX_i) = \bX_i$. However, if the true feature map necessary for linear ignorability to hold also includes higher-order terms or non-linear interactions between covariates, then using only the first-order moments will result in bias \citep{huang2022higher}. As such, omitted variables in such a setting would also include any transformations of existing covariates that have not been explicitly accounted for in the estimated weights. We refer readers to \cite{hartman2022survey} for more discussion about the two assumptions in the context of sensitivity analysis. We note that the proposed sensitivity framework is valid, regardless of which version of conditional ignorability researchers are interested in using.

\subsection{Moving Away from Worst-Case Correlation Bounds} \label{sec:relax_rho}
Theorem \ref{thm:optim_bounds} allows researchers to calculate the maximum bias that can occur for a fixed $R^2$. This is done by assuming the correlation between the imbalance in the omitted confounder is maximally correlated with the outcome. This can be conservative in practice. We provide several recommendations for researchers who may wish to relax this bound. Doing so can result in narrower bounds, at the cost of having to reason about an additional parameter. Throughout this section, we will refer to the correlation bound as $\rho^*$, such that the maximum bias is written as: 
$$\rho^* \cdot \sqrt{\frac{R^2}{1-R^2} \cdot \var(Y_i \mid A_i = 1) \cdot \var(w_i \mid A_i =1)}$$
 We suggest several different approaches for researchers to estimate less conservative bounds. 
 
\paragraph{Estimating Bounds using Relative Correlation} Applying the results from \cite{huang2022sensitivity}, we can decompose the correlation between the imbalance and the outcome into a function of the $R^2$ value, the correlation between the estimated weights and the outcomes, and the correlation between the true weights and the outcomes: 
\begin{equation} 
\cor(w_i, Y_i) \sqrt{\frac{1-R^2}{R^2}} - \cor(w_i^*, Y_i) \cdot \sqrt{\frac{1}{R^2}}
\label{eqn:cor_decomp} 
\end{equation}
As such, an intuitive way to evaluate bounds for the correlation term is to posit a bound for the correlation between the true weights and the outcomes by a relative scaling constant $k$:
$$k := \frac{\cor(w^*_i, Y_i)}{\cor(w_i, Y_i)},$$
where $k$ represents how many more times correlated the true weights are to the outcomes, relative to the estimated weights. $k$ will be naturally upper-bounded at $1/\cor(w_i, Y_i)$. Using Equation \eqref{eqn:cor_decomp}, researchers can then obtain a new upper bound for $\rho^*$: 
$$\rho^* \leq \frac{\cor(w_i, Y_i)}{\sqrt{R^2}} \left(\sqrt{1-R^2} - k\right)$$
It is worth noting that the correlation bound will change, depending on the $R^2$ parameter. 

\paragraph{Benchmarking the Correlation Term} 
 In practice, researchers may also perform formal benchmarking to estimate what may be plausible correlation values. More specifically, researchers can calculate the error from omitting the $j$-th covariate and evaluate the correlation between the residual imbalance in the $j$-th covariate and the outcome, using this as the upper bound for $\rho^*$:
 $$\rho^*_{(j)} \leq \widehat{\cor}(w_i - w_i^{-(j)}, Y_i \mid A_i = 1)$$
Evaluating the bias at $\rho^*_{(j)}$ and $\hat R^2_{(j)}$ provides researchers with an estimate of the bias if they omitted a confounder with residual imbalance that is (1) equivalent in magnitude as the residual imbalance of the $j$-th covariate, and (2) equivalently as correlated with the outcome as the residual imbalance of the $j$-th covariate. Researchers can then estimate the associated confidence intervals by fixing both the correlation term and $R^2$.

\subsection{Extended discussion for sample boundedness}\label{app:overlap} 

\begin{proposition}[Necessary Condition for Validity of Sample Bounds] \label{prop:cond_sample_bounds} \mbox{}\\
Define $\mathcal{A}$ as the set of all observed $Y_i$ values across the sample $A_i = 1$. For sample boundedness to be true (i.e., $\E(Y_i \mid A_i = 0) \in [\min_{i:A_i=1} Y_i, \max_{i:A_i = 1}Y_i])$, the expectation of the outcomes not contained in the sample range must be constrained by the following: 
\begin{align*} 
\E(Y_i &\mid A_i = 0, Y_i \not \in  \mathcal{A})  \in 
&\left[\frac{1}{1-p_\mathcal{A}} \min_{i:A_i =1} Y_i - \frac{p_\mathcal{A}}{1-p_\mathcal{A}} \max_{i:A_i=1} Y _i, \  \frac{1}{1-p_\mathcal{A}} \max_{i:A_i =1} Y_i - \frac{p_\mathcal{A}}{1-p_\mathcal{A}} \min_{i:A_i = 1} Y_i\right],
%\label{eqn:sample_bound_validity}
\end{align*} 
where $p_\mathcal{A} := P(Y_i \in \mathcal{A} \mid A_i = 0)$ represents the proportion of unobserved outcomes that fall within the observed sample range.
\end{proposition} 
The bound specified above represents how much overlap there must exist in the observed and unobserved potential outcomes. The bound is a function of (1) the proportion of unobserved units with outcomes that are outside the range of outcomes across the observed sample units (i.e., $1-p_{\mathcal{A}} = P(Y_i \not \in \mathcal{A} \mid A_i = 0)$), and (2) the sample bounds. If a small proportion of the outcomes in the unobserved population fall outside the sample bounds, then the bound  will be relatively wide. However, if a large proportion of outcomes in the unobserved population fall outside the sample bounds, then the bound will be more narrow.

We also simulate the behavior of both sensitivity models under varying amounts of overlap. 

\begin{example}[Coverage Rates in Limited Outcome Overlap Settings]\mbox{} \label{ex:coverage}\\

Define the treatment assignment mechanism as a logit model, and the outcome model as a linear model: 
$$P(Z_i = 1 \mid \mathcal{X}) \propto \frac{\exp(\gamma_1 X_{i,1} + \gamma_2 X_{i,2} + \beta U_i)}{1+\exp(\gamma_1 X_{i,1} + \gamma_2 X_{i,2}+\beta U_i)} \ \ \ \ \ \ \ \ 
Y_i = \gamma_1 X_{i,1} + \gamma_2 X_{i,2} + \beta U_i + v_i,$$
where $X_{i,1}, X_{i,2}$ and $U_i$ are standard normal random variables, and $v_i \sim N(0, \sigma_v^2)$. $v_i$ represents a noise parameter that controls for how much outcome overlap there is. When $\sigma_v^2$ is large, then there is increased overlap between the treatment and control groups, as the treatment probability is less correlated with the outcome.

We vary $\sigma^2_v \in \{0, 0.1, 0.25, 1, 2, 2.5\}$, and set $\gamma_1 = 2.5$, $\gamma_2 = 5$, and $\beta = 1$. For each iteration of the simulation, we assume that researchers omit $U_i$, and estimate confidence intervals using both the marginal sensitivity model and the variance-based sensitivity model, using the true sensitivity parameters. We visualize the coverage rates across simulations in Figure \ref{fig:coverage}. We see that even in low overlap scenarios and small sample sizes, the variance-based sensitivity model have nominal coverage. However, the marginal sensitivity model struggles to achieve nominal coverage in limited overlap settings.

\end{example} 

Example \ref{ex:coverage} highlights that in small sample settings and limited overlap, the marginal sensitivity model fails to obtain nominal coverage, \textit{even with the true $\Lambda$ value}. In contrast, the variance-based sensitivity model consistently has nominal coverage. 

We see that within finite-sample settings, the marginal sensitivity model may obtain narrower bounds than the variance-based sensitivity model, due to their inherent sample boundedness. However, these narrower bounds risk not being valid in settings with smaller sample size and limited outcome overlap, and can risk large amounts of under-coverage. Thus, the estimated confidence intervals under the variance-based sensitivity model are technically wider, but appropriately so, providing at least nominal coverage, even in cases with severely limited outcome overlap.

\section{Proofs and Derivations} \label{app:proofs}
\subsection{Theorem \ref{thm:optim_bounds}}
\noindent\fbox{%
\vspace{2mm}
\parbox{\textwidth}{%
\vspace{2mm}
For a fixed $R^2 \in [0,1)$, then the maximum bias under $\sigma(R^2)$ can be written as a function of the following components: 
\begin{align*} 
\max_{\tilde w \in \sigma(R^2)} \text{ Bias}(\hat \tau_W \mid \tilde w)
&= \underbrace{\sqrt{1-\cor(w_i, Y_i \mid A_i = 1)^2}}_{\text{(a) Correlation Bound}} \usqrt{\ubrace{\frac{R^2}{1-R^2}}{\makebox[0pt]{\scriptsize\text{(b) Imbalance}}} \cdot \ubrace{\var(Y_i|A_i = 1) \var(w_i | A_i = 1)}{\makebox[0pt]\scriptsize\text{(c) Scaling Factor}}},
\end{align*} 
with the minimum bias given as the negative of Equation \eqref{eqn:optim_bounds}. The optimal bias bounds are thus given by the minimum and maximum biases.
}
}
\begin{proof} 
We will start by deriving the optimal bounds. To begin, we can decompose the bias of a weighted estimator as follows: 
\begin{align} 
\text{Bias}(\hat \tau_W) &= \E \left( \hat \tau_W \right) - \tau \nonumber \\
&\text{By conditional ignorability:} \nonumber \\
&= \E \left( \sum_{i \in \mathcal{A}} w_i Y_i \right) - \E \left( \sum_{i \in \mathcal{A}} w_i^* Y_i\right) \nonumber \\ 
&= \E(w_i Y_i \mid A_i = 1) - \E(w_i^* Y_i \mid A_i = 1) \nonumber \\
&= \E((w_i - w_i^*) \cdot Y_i \mid A_i = 1) \nonumber \\
&\text{By construction, $\E(w_i \mid A_i = 1) = \E(w_i^*\mid A_i = 1)$:} \nonumber \\
&= \E((w_i - w_i^*) \cdot Y_i \mid A_i = 1) - \E(w_i - w_i^* \mid A_i = 1) \cdot \E(Y_i \mid A_i = 1) \nonumber \\
&=\cov(w_i - w_i^*, Y_i \mid A_i = 1) \nonumber \\
&= \cor(w_i - w_i^*, Y_i \mid A_i = 1) \cdot \sqrt{\var(w_i - w_i^* \mid A_i = 1) \cdot \var(Y_i \mid A_i = 1)}
\label{eqn:bias_1} 
\end{align} 
This is similar to the derivation provided in \cite{shen2011sensitivity} and \cite{hong2021did}. However, we will go a step further to amplify the term, $\var(w_i - w_i^* \mid A_i = 1)$, into an $R^2$ value and the variance of the estimated weights. To do so, we extend the results from \cite{huang2022sensitivity}, which examined the bias in the context of an external validity setting, and thus, focused on re-weighting an individual-level treatment effect $\tau_i$. We instead apply the results to a general missingness setting, in which we are re-weighting outcomes $Y_i$. We re-write the variance of the error in the weights in Equation \eqref{eqn:bias_1} as a function of the $R^2$ parameter and the variance of the estimated weights, providing the following bias decomposition:
$$\text{Bias}(\hat \tau_W) = \cor(w_i - w_i^*, Y_i \mid A_i = 1) \cdot \sqrt{\frac{R^2}{1-R^2} \cdot \var(Y_i \mid A_i = 1) \cdot \var(w_i \mid A_i = 1)},$$
where $R^2$ is defined in Definition \ref{def:vbm}. Because we are fixing $R^2 \in [0,1)$,\footnote{In settings when $R^2 = 1$, this implies that researchers have effectively explained \textit{none} of the variation in the true weights--i.e., in settings when researchers use uniform weights. However, if researchers have at least included one covariate that is at least correlated with a variable in the separating set $\mathcal{X}$, then $R^2 < 1$.} and $\var(Y_i \mid A_i = 1) \cdot \var(w_i \mid A_i = 1)$ are directly estimable from the data, to maximize the bias, we must maximize the correlation term. 

Applying results from \cite{huang2022sensitivity}, we note that the error in the weights (i.e., $w_i - w_i^*$) is orthogonal to the estimated weights $w_i$ (i.e., $\cov(w_i - w_i^*, w_i \mid A_i = 1) = 0$). Then, applying the recursive formula of partial correlation, we obtain the following bounds for the correlation:\footnote{This follows from results in \cite{olkin1981range}, which shows that for any three random vectors $a$, $b$, and $c$, the correlation of $a$ and $c$ can be bounded by: 
$$\cor(a,b) \cor(c, b) -\sqrt{1-\cor^2(a, b)}\cdot \sqrt{1-\cor^2(b,c)} \leq \cor(a,c) \leq \cor(a,b) \cor(c, b) + \sqrt{1-\cor^2(a, b)}\cdot \sqrt{1-\cor^2(b,c)}.$$

}
$$-\sqrt{1-\cor(w_i, Y_i \mid A_i = 1)^2}\leq \cor(w_i - w_i^*, Y_i \mid A_i = 1) \leq \sqrt{1-\cor(w_i, Y_i \mid A_i = 1)^2}$$
Thus, Equation \ref{eqn:optim_bounds} in Theorem \ref{thm:optim_bounds} directly follows. 

\end{proof}

%%%%%%%%%%%%%%%%%%%%%%%%%%%%%%%%%%%%%%%%%%%%%%%%%%%%%%%%%%%%%%%%%%%%%%%%%%%%%%%%%%%%%%%%%%%%%%%%%%%%%%%%%%%%%%%%
\subsection{Theorem \ref{thm:bs}} 
\noindent\fbox{%
\vspace{2mm}
\parbox{\textwidth}{%
\vspace{2mm}
For every $\tilde w \in \sigma(R^2)$: 
$$\limsup_{n \to \infty} P(\tau(\tilde w) < L(\tilde w)) \leq \frac{\alpha}{2} \text{ and } \limsup_{n \to \infty} P(\tau(\tilde w) > U(\tilde w)) \leq \frac{\alpha}{2},$$
where $L(\tilde w)$ and $U(\tilde w)$ are defined as the $\alpha/2$ and $1-\alpha/2$-th quantiles of the bootstrapped estimates (i.e., Equation \eqref{eqn:def_bounds}). 
}
}
\begin{proof} 
We may re-write our bootstrapped estimate $\hat \tau^{(b)}(\tilde w)$ as: 
\begin{align*} 
\hat \tau^{(b)}(\tilde w) &= \hat \tau_W^{(b)} - \text{Bias}(\hat \tau_W \mid \tilde w)\\
&= \hat \tau_W^{(b)} - \rho \cdot \sqrt{\var(\hat w_i^{(b)}) \frac{R^2}{1-R^2} \cdot \var(Y_i^{(b)})}
\end{align*} 
Because $\rho$ and $R^2$ are fixed (across bootstrap samples), the components that drive variation across bootstrap samples are: $\hat \tau_W^{(b)}$, $\var(\hat w_i^{(b)})$, and $\var(Y_i^{(b)})$. 

An overview of the proof is as follows. Similar to \cite{zhao2019sensitivity}, we will use a $Z$-estimation framework. In particular, we will add in three additional parameters: $\hat \mu_w^2$, $\hat \mu_Y$, $\hat \mu_Y^2$, which represent the second order moment of the weights, the average of the outcomes, and the second order moment of the outcomes, respectively. Then, we will invoke the asymptotic normality of bootstrapped $Z$-estimators. In the following proof, we will show the validity of the percentile bootstrap in the case that researchers are using inverse propensity score weights; however, we note that researchers can invoke the results in \cite{soriano2021interpretable} to show validity of the results for balancing weights. 

To begin, define $\mu_w$ as the expectation of the weights: 
$$\mu_w = \E(A w) \equiv \E(A \cdot (1+\exp(-\beta X))).$$ 
Then, we define $\mu$ as:  
$$\mu = \frac{\E(AY(1+\exp(-\beta^\top X)))}{\mu_w}.$$
Define $\mu^2_w = \E(A w^2)$ and $\sigma^2_Y = \E(A Y^2)$ as the second moment of the weights and the outcomes, respectively. Then, we define the vector $\theta = \left(\mu, \mu_w, \beta, \mu^2_w, \mu_Y, \mu^2_Y \right)^\top \in \Theta$. Define the function $Q:{0,1} \times \R^d \times \R \to \R^{d+5}$, where for $t = (a, x^\top, y) \in \{0,1\} \times \R^d \times \R$: 
\begin{align} 
Q( t\mid \theta) = \begin{pmatrix}
Q_1(t|\theta)  \\ 
Q_2(t|\theta)  \\
Q_3(t|\theta)  \\
Q_4(t|\theta)  \\
Q_5(t|\theta)  \\
Q_6(t|\theta)  
\end{pmatrix} := 
\begin{pmatrix} 
\left( a - \frac{\exp(\beta^\top x)}{1+\exp(\beta^\top x)} \right) x \\
 \mu_w - a\left( 1+\exp(-\beta^\top x) \right) \\
 \mu_w \mu - ay \left(1+ \exp(-\beta^\top x) \right) \\
 \mu_w^2 - a \left(1+\exp(-\beta^\top x)\right)^2\\ 
 \mu_y - ay \\
 \mu_y^2 - ay^2 
\end{pmatrix} 
\end{align} 
Finally, we define $\Phi(\theta)$ as: 
$$\Phi(\theta) = \int Q(t | \theta) d \mathbb{P}(t),$$
where $T = (A, X^\top, AY)^\top \sim \mathbb{P}$, where $\mathbb{P}$ represents the true distribution generating the data. It is simple to see that $\Phi(\theta^*) = 0$, when $\theta^*$ is equal to the true parameter values. Then, the $Z$-estimates $\hat \theta$ : 
\begin{align} 
\Phi_n(\hat \theta) :&= \frac{1}{n} \sum_{i=1}^n Q(T_i | \hat \theta) \\
&= \begin{pmatrix} 
\left( \frac{1}{n} \sum_{i=1}^n A_i - \frac{\exp(\hat \beta^\top \bX_i)}{1+\exp(\hat \beta^\top \bX_i)} \right) \bX_i \\
 \hat \mu_w - \frac{1}{n} \sum_{i=1}^nA_i\left( 1+\exp(-\hat \beta^\top \bX_i) \right) \\
 \hat \mu_w \mu - \frac{1}{n} \sum_{i=1}^nA_i Y_i \left(1+ \exp(-\hat \beta^\top \bX_i) \right) \\
 \hat \mu^2_w - \frac{1}{n} \sum_{i=1}^n A_i \left(1+\exp(-\hat \beta^\top \bX_i)\right)^2\\ 
 \hat \mu_y - \frac{1}{n} \sum_{i=1}^n A_i Y_i \\
 \hat \mu_y^2 - \frac{1}{n} \sum_{i=1}^n ( A_i Y_i^2)
\end{pmatrix} = 0 
\end{align} 
We define $\Sigma := \E(Q(t \mid \theta)Q(t \mid \theta)^\top)$.
We will invoke the following regularity conditions, consistent with \cite{zhao2019sensitivity}. 
\begin{assumption}[Regularity Conditions] \label{assump:regularity_conds} \mbox{}\\
Assume that the parameter space $\Theta$ is compact, and that $\theta$ is in the interior of $\Theta$. Furthermore, $(Y, \bX)$ satisfies the following: 
\begin{enumerate} 
\item $\E(Y^4) <\infty$
\item $\det \left( \E \left( \frac{\exp(\beta^\top \bX)}{(1+\exp(\beta^\top \bX))^2} \bX \bX^\top \right) \right) > 0$
\item $\forall$ compact subsets $S \subset \R^d$, $\E(\sup_{\beta \in S} \exp(\beta^\top \bX) ) < \infty$
\end{enumerate} 
\end{assumption} 

To show asymptotic normality of bootstrapped $Z$-estimators, we must first verify that $\dot{\Phi}_0$ and $\Sigma$ are well-defined. 
\begin{align*} 
\dot{\Phi}_0 &= \E \left(\nabla_{\theta = \theta_0} Q(T | \theta) \right)\\
&=\begin{pmatrix} 
0 & 0 & -\E \left( \frac{\exp(\beta_0^\top \bX)}{1 + \exp(\beta_0^\top \bX)^2} \bX \bX^\top \right) & 0 & 0 & 0 \\
0 & 1 & \E(A \bX^\top \exp(-\beta_0^\top \bX) & 0 & 0 & 0\\
\mu_w & \mu & \E(A Y \bX^\top (\exp(\beta_0^\top \bX) & 0 & 0 & 0\\ 
0 & 0 & \E(A \bX^\top (\exp(\beta_0^\top \bX) +\exp(-2\beta_0^\top \bX)) ) & 1 & 0 & 0\\
0 & 0 & 0 & 0 & 1 & 0\\
0 & 0 & 0 & 0 & 0 & 1 
\end{pmatrix} 
\end{align*} 

By Leibniz Formula: 
\begin{align*} 
\left|\det(\dot \Phi_0)\right| &= \left|\det \begin{pmatrix} 
0 & 0 & -\E \left( \frac{\exp(\beta_0^\top \bX)}{1 + \exp(\beta_0^\top \bX)^2} \bX \bX^\top \right) \\
0 & 1 & \E(A \bX^\top \exp(-\beta_0^\top \bX) \\
\mu_w & \mu & \E(A Y \bX^\top (\exp(\beta_0^\top \bX) 
\end{pmatrix} \det \begin{pmatrix} 1 & 0 & 0 \\
0 & 1 & 0 \\
0 & 0 & 1\end{pmatrix} \right|\\
&= \mu_w \left| \det \E \left( \frac{\exp(\beta_0^\top \bX)}{(1+\exp(\beta_0^\top \bX))^2} \bX \bX^\top \right)\right| > 0,
\end{align*} 
which follows by regularity condition (2). As such, $\dot{\Phi}_0$ is invertible. Furthermore, by regularity condition (1), $\Sigma < \infty$.

As such, we simply need to verify the three conditions for asymptotic normality of bootstrapped $Z$-estimators: 
\begin{enumerate} 
\item The class of functions ${t \to Q(t | \theta): \theta \in \Theta}$ is $\mathbb{P}$-Glivenko-Cantelli. 
\item $||\Phi (\theta)||_1$ is strictly positive outside every open neighborhood of $\theta_0$.
\item The class of functions is $\mathbb{P}$-Donsker, and $\E((Q(T|\theta_n) - Q(T|\theta_0))^2) \to 0$ whenever $||\theta_n - \theta_0||_1 \to 0$.
\end{enumerate} 
It is worth noting that the first three parameters ($\mu$, $\mu_w$, $\beta$) are special cases from \cite{zhao2019sensitivity}, in which we do not perform any shifting in the weights (i.e., $h(x, y) = 0$). We will then show that the three conditions still hold after additionally accounting for the last three parameters. The proof for each condition is provided below. \\

\noindent \textbf{Condition 1:} The class of functions ${t \to Q(t | \theta): \theta \in \Theta}$ is $\mathbb{P}$-Glivenko-Cantelli. 
\begin{align*} 
|| Q(t | \theta) \leq || Q_1(t|\theta)||_1 + \sum_{b=2}^5 |Q_b(t| \theta)|
\end{align*}  
\cite{zhao2019sensitivity} show that $||Q_1(t|\theta)||_1 + |Q_2(t|\theta)| + |Q_3(t|v)|$ is bounded as a function of $x$, $y$, and some absolute constant $M_1$:
$$||Q_1(t|\theta)||_1 + |Q_2(t|\theta)| + |Q_3(t|v)| \leq || x ||_1 + |y| + \exp(-\beta^\top x)(1+|y|) + M_1.$$
As such, all that is left to show is to show that $|Q_4(t|\theta)| + |Q_5(t|\theta)|+|Q_6(t|\theta)|$ is finite. To begin: 
\begin{align*} 
|Q_4(t|\theta)| &= |\mu^2_w - (a(1+\exp(-\beta^\top x)^2|\\
&\leq \mu_w^2 + (1+\exp(-\beta^\top x))^2 \\
|Q_5(t|\theta)| &= |\mu_y - ay|\\
&\leq |\mu_y| + |y| \\
|Q_6(t|\theta)| &= |\mu^2_y - ay^2|\\
&\leq \mu^2_y + |y^2|
\end{align*} 
As such: 
\begin{align*} 
|Q_4(t|\theta)| + |Q_5(t|\theta)|+ |Q_6(t|\theta)| \leq M_2 + (1+\exp(-\beta^\top x))^2 + |y| + |y^2|,
\end{align*} 
where $M_2$ is some absolute constant. As such, where $M$ is an absolute constant:
$$||Q(t| \theta ||_1 \leq ||x||_1 + 2|y| + |y^2| +\exp(-\beta^\top x)(1+|y|) + (1+\exp(-\beta^\top x))^2 + M,$$ 
where $M < \infty$ by regularity condition (1). 
%\sam{Remind us which regularity conditions ensure this RHS is finite.}
Therefore, $\E(\sup_{\theta \in \Theta}||Q(t|\theta)||_1) < \infty$, and $\{t \to Q(t|\theta):\theta \in \Theta\}$ is $\mathbb{P}$-Glivenko-Cantelli.\\

\noindent \textbf{Condition 2:} $||\Phi (\theta)||_1$ is strictly positive outside every open neighborhood of $\theta_0$.\\
Following \cite{zhao2019sensitivity}, we fix some $\varepsilon > 0$. If $|| \beta - \beta_0||_1 > \varepsilon/M$, then it is trivial to show that $|| \Phi(\theta)||_1 > 0$. \cite{zhao2019sensitivity} show that when $||\beta-\beta_0||_1 \leq \varepsilon/M$, if $|\mu_w - \mu_{w,0}| > \varepsilon/4K$, where $K = \sup_{\theta \in \Theta} |\mu| \in (0,\infty)$, 
then $|| \Phi(\theta)||_1 > 0$. Furthermore, when $||\beta-\beta_0||_1 \leq \varepsilon/M$ and $|\mu_w - \mu_{w,0}| \leq \varepsilon/4K$ and $|\mu - \mu_0| > \varepsilon/2\mu_w$, then $||\Phi(\theta)||_1 > 0$. 

Thus, we must show for the remaining 3 parameters that when $||\mu^2_w - \mu^2_{w,0}||$, $||\mu_y - \mu_{y,0}||$, or $||\mu^2_y - \mu^2_{y,0}||_1$ are greater than some $\varepsilon$, $|| \Phi(\theta)||_1 > 0$. Assume $||| \beta - \beta_0 ||_1 \leq \varepsilon/M$. Then: 
\begin{align*} 
\left| \E \left[A \exp(-\beta^\top \bX)^2 + A \exp(-\beta_0^\top \bX)^2 \right] \right|
&=\left| \E \left[ A \exp(-2\beta^\top \bX) + A \exp(-2 \beta_0^\top \bX) \right] \right|\\
&\leq \left|\E \left( \exp(-2\beta^\top \bX - 2 \beta_0^\top \bX) \right) \right| \\
&\leq 2 || \beta - \beta_0||_\infty \E \left( || \bX||_1 \exp(-t^*) \right) \text{ for }t^* \in [\beta_0^\top \bX, \beta^\top \bX]\\
&\leq 2 \cdot \frac{\varepsilon}{64K} = \frac{\varepsilon}{32K}
\end{align*} 
As such, if $||\mu^2_w - \mu^2_{w,0}|| > \varepsilon/32K$:
\begin{align} 
|| \Phi(\theta)||_1 \geq \left|\mu^2_w - \mu^2_{w,0}  + \E \left[A \exp(-\beta^\top \bX)^2 + A \exp(-\beta_0^\top \bX)^2 \right] \right| > 0
\label{eqn:mu2_cond} 
\end{align} 

For the final two parameters, it is worth noting that there is no dependency on the other parameter estimates. As such, regardless of whether the other parameters are smaller than some $\epsilon$, if $||\mu_y - \mu_{y,0}||_1 >\varepsilon$:
\begin{align} 
||\Phi(\theta)||_1 &\geq \left|\mu_y - \E(AY) - (\mu_{y,0} - \E(AY)) \right| \nonumber \\
&= \left| \mu_y - \mu_{y,0} \right|  > 0 
\label{eqn:muy_cond} 
\end{align} 
Similarly, if $||\mu^2_y - \mu^2_{y,0}|| > \varepsilon$
\begin{align} 
||\Phi(\theta)||_1 &\geq \left|\mu^2_y - \E(AY^2) - (\mu^2_{y,0} - \E(AY^2)) \right| \nonumber \\
&= \left| \mu^2_{y,0} - \mu_{y} \right| > 0 
\label{eqn:muy2_cond} 
\end{align} 

As such, combining Equation \eqref{eqn:mu2_cond}, \eqref{eqn:muy_cond}, \eqref{eqn:muy2_cond}, as well as the results from \cite{zhao2019sensitivity}, we have shown that for all $\delta > 0$, $\inf\{ ||\Phi(\theta)||^2: || \theta - \theta_0||_1 >\delta\} > 0$. \\

\noindent \textbf{Condition 3:} The class of functions is $\mathbb{P}$-Donsker, and $\E((Q(T|\theta_n) - Q(T|\theta_0))^2) \to 0$ whenever $||\theta_n - \theta_0||_1 \to 0$.\\
From \cite{zhao2019sensitivity}, we obtain a bound for the first three terms (i.e., $Q_1(t| \theta)$, $Q_2(t| \theta)$, and $Q_3(t| \theta)$). Then, for the 4th term:\footnote{Consistent with \cite{zhao2019sensitivity}, for some $a,b\in \R$, and some constant $C>0$, if $a \leq C \cdot b$, then we write $a \lesssim b$.}
\begin{align*} 
|Q_4&(t|\theta_2) - Q_4(t|\theta_1) | \\
&= | \mu^2_{w,2} - (a(1+\exp(-\beta_2^\top x))^2 - 
(\mu^2_{w,1} - (a(1+\exp(-\beta_1^\top x))^2|\\
&\leq | \mu^2_{w,2} - \mu^2_{w,1} | + |(1+\exp(-\beta_2^\top x))^2 - (1+\exp(-\beta_1^\top x))^2| \\
&=| \mu^2_{w,2} - \mu^2_{w,1} |+ \left| 2 \big( \exp(-\beta_2^\top x) - \exp(-\beta_1^\top x) \big) + \exp(-2 \beta_2^\top x) - \exp(-2\beta_1^\top x) \right|\\
&\leq | \mu^2_{w,2} - \mu^2_{w,1} |+ \left| 2 \big( \exp(-\beta_2^\top x) - \exp(-\beta_1^\top x) \big) \right| + \left|\exp(-2 \beta_2^\top x) - \exp(-2\beta_1^\top x) \right|
\intertext{Applying the Mean Value Theorem (equivalently, results from \cite{zhao2019sensitivity}):} 
&\lesssim | \mu^2_{w,2} - \mu^2_{w,1} | + 2 || \beta_2 - \beta_1 ||_2 ||x||_2 \sup_{\beta \in \Theta} \exp(-\beta^\top x) + || 2\beta_2 - 2\beta_1 ||_2 ||x||_2 \sup_{\beta \in \Theta } \exp(-2\beta^\top x) \\
&= | \mu^2_{w,2} - \mu^2_{w,1} | + 2 || \beta_2 - \beta_1 ||_2 ||x||_2 \sup_{\beta \in \Theta} \exp(-\beta^\top x) + 2|| \beta_2 - \beta_1 ||_2 ||x||_2 \sup_{\beta \in \Theta } \exp(-2\beta^\top x) \\
&= | \mu^2_{w,2} - \mu^2_{w,1} | + 2|| \beta_2 - \beta_1 ||_2 ||x||_2  \sup_{\beta \in \Theta} \exp(-\beta^\top x) \left(1+ \sup_{\beta \in \Theta } \exp(-\beta^\top x) \right)\\
&\lesssim M_4(x) \left(| \mu^2_{w,2} - \mu^2_{w,1} | + || \beta_2 - \beta_1 ||_1\right)
\end{align*}

\noindent Finally, for the 5th and 6th terms: 
\begin{align*} 
|Q_5&(t|\theta_2) - Q_5(t|\theta_1) | \\
&= | \mu_{y,2} - ay - (\mu_{y,1} - ay)| \\
&= | \mu_{y,2} - \mu_{y,1} | \\
|Q_6&(t|\theta_2) - Q_6(t|\theta_1) | \\
&= | \mu^2_{y,2} - ay^2  - (\mu^2_{y,1} - ay^2) | \\
&\leq | \mu^2_{y,2} - \mu^2_{y,1} | 
\end{align*} 

Combining results with \cite{zhao2019sensitivity}, we see that: 
\begin{align*} 
||Q(t | \theta_2) - Q(t | \theta_1) ||_1 = \sum_{b=1}^6 ||Q_b(t | \theta_2) - Q_b(t | \theta_1)|| \lesssim M(x,y) || \theta_2 - \theta_1||_1 
\end{align*} 
Since $\E(M(X,Y)^2) < \infty$, we have shown that the class of functions is $\mathbb{P}$-Donsker, and furthermore, that whenever $||\theta_n - \theta_0||_1 \to 0$, $\E\big[(Q(t |\theta_n) - Q(t|\theta_0))^2\big] \to 0$. 

Then, by invoking \cite{kosorok2008introduction}, Theorem 10.16:
\begin{equation} 
\sqrt{n} (\hat \theta - \theta) \cid N \left(0, \dot \Phi^{-1}_0 \Sigma \dot \Phi_0 \right), \ \ \text{ and } \ \ \sqrt{n}(\hat \theta^{(b)} - \theta) \cid N \left(0, \dot \Phi^{-1}_0 \Sigma \dot \Phi_0 \right),
\end{equation} 
As such, applying Delta Method and results from Appendix C3 in \cite{zhao2019sensitivity} concludes the proof. 
\end{proof} 
\subsection{Theorem \ref{thm:l2_norm} (Weighted $L_2$ Analog)} 
\noindent\fbox{%
\vspace{2mm}
\parbox{\textwidth}{%
\vspace{2mm}

Define the individual-level error in the weights as $\lambda_i := w_i^*/w_i$. Define the $L_{2,w}$ norm as follows: 
$$||\lambda||^2_{2,w} := \begin{cases} \displaystyle 
\frac{1}{n} \sum_{i=1}^n \lambda_i^2 \cdot \nu(w_i) & \text{if } \var(w_i) > 0, \\
\infty & \text{else}
\end{cases},$$
where $\nu(w_i)$ is a function of the estimated weights. Then, the variance-based sensitivity model can equivalently be written as a norm-constrained optimization problem:
$$\max_{\tilde w \in \sigma(R^2)} \text{Bias}(\hat \tau_W \mid \tilde w) \iff \begin{cases} \displaystyle \max_{\tilde w} \text{Bias}(\hat \tau_W \mid \tilde w) \\ 
\displaystyle \text{s.t. } ||\lambda||_{2,w} \leq \sqrt{\frac{k}{1-R^2}},
\end{cases}$$
where $k := 1-R^2/\E(w_i^2)$.
}
}

\begin{proof} 
Define $\lambda_i := w_i^*/w_i$. Then, following results from \cite{huang2022sensitivity}:  
\begin{align*} 
\var(w_i - w_i^*) &= \var(w_i^*) - \var(w_i) 
\intertext{We can then substitute in $\lambda_i$:}
&= \var(\lambda_i \cdot w_i) - \var(w_i) \\
&= \E(\lambda_i^2 \cdot w_i^2) - \underbrace{\E(\lambda_i \cdot w_i)^2}_{\equiv \E(w_i^*)^2 = 1} - \var(w_i) \\
&= \cov(\lambda_i^2, w_i^2) + \E(\lambda_i^2)\E(w_i^2) - 1 - \var(w_i)\\
&= \cov(\lambda_i^2, w_i^2) + \E(\lambda_i^2) \var(w_i) + \E(\lambda_i^2) - 1 - \var(w_i) \\
&= \cov(\lambda_i^2, w_i^2) + (\E(\lambda_i^2) - 1) \cdot \left( \var(w_i) + 1 \right)\\
&= \cov(\lambda_i^2, w_i^2) + (\E(\lambda_i^2) - 1) \cdot \E(w_i^2)\\
\implies \frac{\var(w_i-w_i^*)}{\E(w_i^2)} &= \frac{\cov(\lambda_i^2, w_i^2)}{\E(w_i^2)} + (\E(\lambda_i^2) - 1)
\end{align*} 
Re-arranging the terms: 
\begin{align*} 
\frac{\cov(\lambda_i^2, w_i^2)}{\E(w_i^2)} + \E(\lambda_i^2) &= 1 + \frac{\var(w_i - w_i^*)}{\E(w_i^2)} \\
&= 1 + \frac{\E(w_i^2) - \E(w_i)^2}{\E(w_i^2)} \cdot \frac{R^2}{1-R^2}\\
&= 1 + \frac{R^2}{1-R^2} - \frac{\E(w_i)^2}{\E(w_i^2)} \cdot \frac{R^2}{1-R^2}\\
&= \frac{1}{1-R^2}- \underbrace{\frac{\E(w_i)^2}{\E(w_i^2)}}_{1/\E(w_i^2)} \cdot \frac{R^2}{1-R^2}\\
&= \frac{1}{1-R^2} \underbrace{\left( 1-  \frac{R^2}{\E(w_i^2)} \right)}_{:= k}
\end{align*} 
By setting $R^2$, we are also setting the value for $\frac{\cov(\lambda_i^2, w_i^2)}{\E(w^2_i)} + \E(\lambda_i^2)$. 

We now re-write $\frac{\cov(\lambda_i^2, w_i^2)}{\E(w^2_i)} + \E(\lambda_i^2)$ as a weighted sum:  
\begin{align*} 
\E(\lambda_i^2) + \frac{\cov(\lambda_i^2, w_i^2)}{\E(w_i^2)} &= \frac{1}{n} \sum_{i=1}^n \lambda_i^2 + \frac{1}{\E(w_i^2)} \cdot \frac{1}{n} \sum_{i=1}^n (\lambda_i^2 - \E(\lambda_i^2)) (w_i^2 - \E(w_i^2))\\
&= \frac{1}{n} \sum_{i=1}^n \lambda_i^2 + \frac{1}{n} \sum_{i=1}^n \lambda_i^2 \cdot \frac{w_i^2 - \E(w_i^2)}{\E(w_i^2)} - \frac{1}{n} \sum_{i=1}^n \E(\lambda_i^2) \cdot \frac{w_i^2 - \E(w_i^2)}{\E(w_i^2)} \\
&= \frac{1}{n} \sum_{i=1}^n \lambda_i^2 \cdot \underbrace{\left( 1 +\frac{w_i^2 - \E(w_i^2)}{\E(w_i^2)}  \right)}_{:=\nu(w_i)} + \E(\lambda_i^2) \underbrace{\frac{1}{n}\sum_{i=1}^n \frac{w_i^2 - \E(w_i^2)}{\E(w^2_i)} }_{:= 0} \\
&= \frac{1}{n} \sum_{i=1}^n \lambda_i^2 \cdot \nu(w_i)
\end{align*} 

As such, we can define the $L_{2,w}$ norm as follows: 
$$|| \lambda ||^2_{2,w} := \begin{cases}
\frac{1}{n} \sum_{i=1}^n \lambda_i^2 \cdot \nu(w_i) & \text{if } \var(w_i) > 0 \\
\infty & \text{else}
\end{cases} $$

We will show that $L_{2,w}$ meets the criteria for being a semi-norm. 

\begin{enumerate} 
\item Triangle Inequality: $||\lambda_1 + \lambda_2 ||_{2,w} \leq || \lambda_1 ||_{2,w} +   || \lambda_2 ||_{2,w} $
\begin{align*} 
||\lambda_1 + \lambda_2 ||^2_{2,w} &= \sum_{i=1}^n (\lambda_{i,1} + \lambda_{i,2})^2 \cdot \nu(w_i)  \\
&= \sum_{i=1}^n \lambda^2_{i,1} \nu(w_i) + \sum_{i=1}^n\lambda^2_{i,2} \nu(w_i) + 2 \sum_{i=1}^n\lambda_{i,1} \lambda_{i,2} \nu(w_i)
\intertext{Applying Cauchy-Schwarz: $\left(\sum_{i=1}^n\lambda_{i,1} \lambda_{i,2} \nu(w_i)\right)^2 \leq \left(\sum_{i=1}^n\lambda_{i,1} \lambda_{i,1}^2 \nu(w_i) \right)^2 \left(\sum_{i=1}^n\lambda_{i,1}\lambda_{i,2}^2 \nu(w_i)\right)^2$. Then:} 
&\leq \sum_{i=1}^n \lambda^2_{i,1} \nu(w_i) + \sum_{i=1}^n\lambda^2_{i,2} \nu(w_i) + 2 \left(\sum_{i=1}^n\lambda_{i,1} \lambda_{i,1}^2 \nu(w_i) \right) \left(\sum_{i=1}^n\lambda_{i,1}\lambda_{i,2}^2 \nu(w_i)\right)\\
&= \left(||\lambda_1||_{2,w} + ||\lambda_2||_{2,w}\right)^2
\end{align*} 
\item Absolute homogeneity: 
\begin{align*} 
||k \cdot \lambda||_{2,w} &= \sqrt{\sum_{i=1}^n (k \cdot \lambda_i)^2 \cdot \nu(w_i) } \\
&= k \sqrt{\sum_{i=1}^n \lambda_i^2 \cdot \nu(w_i) }\\
&= k \cdot || \lambda||_{2,w}
\end{align*} 
\end{enumerate} 
\end{proof} 
\subsection{Theorem \ref{thm:bound_cond}} 
\noindent\fbox{%
\vspace{2mm}
\parbox{\textwidth}{%
\vspace{2mm}
Let $\psi(\Lambda)$ represent the difference in the estimated point estimate bounds under the marginal sensitivity model $\varepsilon(\Lambda)$ for a given $\Lambda \geq 1$: 
$$\psi(\Lambda):= \max_{\tilde w \in \varepsilon(\Lambda)} \frac{\sum_{i:Z_i = 0} Y_i Z_i \tilde w_i}{\sum_{i:Z_i=0} Z_i \tilde w_i}- \min_{\tilde w \in \varepsilon(\Lambda)}\frac{\sum_{i:Z_i = 0} Y_i Z_i \tilde w_i}{\sum_{i:Z_i=0} Z_i \tilde w_i}.$$ 
Then if the true $R^2$ parameter is lower than the following threshold,
\begin{equation*} 
R^2 \leq \frac{\psi(\Lambda)^2}{4 \underbrace{(1-\cor(w_i, Y_i)^2)}_{\text{Correlation Bound}} \cdot \underbrace{\var(w_i) \var(Y_i)}_{\text{Scaling Factor}} + \psi(\Lambda)^2},
\end{equation*} 
the bounds under the variance-based sensitivity model will be narrower than the bounds for the marginal sensitivity model.
}
}
\begin{proof} 
The length of the point estimate bounds under the variance-based sensitivity model $\sigma(R^2)$ is equal to two times the maximum bias bound: 
\begin{align*} 
\max_{\tilde w \in \sigma(R^2)}  \tau(\tilde w) - \min_{\tilde w \in \sigma(R^2)}  \tau(\tilde w) =
2 \cdot \sqrt{1-\cor(w_i, Y_i \mid A_i = 1)^2} \cdot \sqrt{\frac{R^2}{1-R^2} \cdot \var(w_i) \cdot \var(Y_i \mid A_i = 1)}
\end{align*} 
By definition, the length of the estimated point estimate bounds under the marginal sensitivity model is represented by $\psi(\Lambda)$.  Thus, we want to solve for the $R^2$ value such that the following inequality holds: 
\begin{align*} 
2 \cdot \sqrt{1-\cor(w_i, Y_i \mid A_i = 1)^2} \cdot \sqrt{\frac{R^2}{1-R^2} \cdot \var(w_i \mid A_i = 1) \cdot \var(Y_i \mid A_i = 1)} \leq \psi(\Lambda) 
\end{align*} 
Solving for the $R^2$ value: 
\begin{align*} 
\frac{R^2}{1-R^2} &\leq \frac{\psi(\Lambda)^2/4}{(1-\cor(w_i, Y_i \mid A_i = 1)^2) \cdot \var(w_i \mid A_i = 1) \cdot \var(Y_i \mid A_i = 1)} \\
R^2 &\leq \frac{\psi(\Lambda)^2/4 (1-\cor(w_i, Y_i \mid A_i = 1)^2) \var(w_i \mid A_i = 1) \var(Y_i \mid A_i = 1)}{1+\psi(\Lambda)^2/4 (1-\cor(w_i, Y_i \mid A_i = 1)^2) \var(w_i \mid A_i = 1) \var(Y_i \mid A_i = 1)}\\
&= \frac{\psi(\Lambda)^2}{4 (1-\cor(w_i, Y_i \mid A_i = 1)^2) \var(w_i \mid A_i = 1) \var(Y_i \mid A_i = 1) + \psi(\Lambda)^2},
\end{align*} 
The results from the corollary directly follow.
\end{proof} 
\subsection{Example \ref{ex:infty_lambda}}
\noindent\fbox{%
\vspace{2mm}
\parbox{\textwidth}{%
\vspace{2mm}
Assume researchers use a logit model to estimate the weights using $\bX_i$, but omit a confounder $U_i$. The estimated and ideal weights take on the following forms: 
\begin{equation*} 
\hat w_i = \exp(\hat \gamma^\top \bX_i) \ \ \ \ \ \ \ \ \ \ \hat w_i^* = \exp(\hat \gamma^{*\top} \bX_i + \hat \beta U_i)
\end{equation*} 
Assume $\begin{bmatrix} \bX_i,U_i\end{bmatrix} \iid MVN(0, I)$. Then $\E(\hat \Lambda) \to \infty$ as $n \to \infty$: 
$$\lim_{n \to \infty} \frac{\E(\hat \Lambda)}{\exp(\sqrt{2 \nu^2 \log(n)})} \geq 1,$$
where $\nu^2 = (\gamma^* - \gamma)^\top (\gamma^* - \gamma) + \beta^2$, and the results follow immediately from \cite{wainwright2019high}. 
}
}
\begin{proof} 
The multiplicative error between $w_i^*$ and $w_i$ is written as: 
\begin{align*} 
\frac{\hat w_i^*}{\hat w_i} = \frac{\exp(\hat \gamma^{*\top} \bX_i + \hat \beta U_i)}{\exp(\hat \gamma^\top \bX_i)} = \exp((\hat \gamma^{*} - \hat \gamma)^\top \bX_i +\hat \beta U_i),
\end{align*} 
and $\hat \Lambda$ is defined as the maximum: 
$$\hat \Lambda = \max_{1 \leq i \leq n} \exp(|(\hat \gamma^{*} - \hat \gamma)^\top \bX_i + \hat \beta U_i|)$$
We will show that $\E(\hat \Lambda) \to \infty$, as $n \to \infty$. To begin, define $V_i$ as: 
$$ V_i := (\hat \gamma^* - \hat \gamma)^\top \bX_i + \hat \beta U_i$$ 
Because $\bX_i$ and $U_i$ are normally distributed, $V_i$ will be normally distributed, with mean 0, and variance $\nu^2:=(\hat \gamma^* - \hat \gamma)^\top + \hat \beta^2$.
Let $V^{(1)}, ..., V^{(n)}$ be the ordered set of $V$ such that $V^{(1)} \leq ... \leq V^{(n)}$. Without loss of generality, assume $|V^{(n)}| \geq | V^{(1)}|$. Then, $\E(\hat \Lambda) = \E(\exp(| V^{(n)}|)$. Using Jensen's inequality, the expectation of $\hat \Lambda$ may be lower bounded: 
$$\E(\hat \Lambda) = \E(\exp(|V^{(n)}|) \geq \exp(\E(|V^{(n)}|))$$
Then, we may invoke a well-studied result that for any set of $n$ normally distributed random variables (\cite{wainwright2019high}): 
$$\lim_{n \to \infty} \frac{\E(V^{(n)})}{\sqrt{2 \nu^2 \log(n)}} = 1$$
Because $\E(|V^{(n)}|) \geq \E(V^{(n)})$
$$\lim_{n \to \infty} \frac{\E(V^{(n)})}{\sqrt{2 \nu^2 \log(n)}} = 1$$
As such, as $n \to \infty$, $\E(\hat \Lambda) \to \infty$.
\end{proof} 

\subsection{Example \ref{ex:r2_logit}}
\noindent\fbox{%
\vspace{2mm}
\parbox{\textwidth}{%
\vspace{2mm}
Consider the same setting as Example \ref{ex:infty_lambda}. Then, the $R^2$ value can be written as follows: 
$$R^2 = 1 - \frac{\exp(\hat \gamma^\top \hat \gamma) - 1}{\exp(\hat \gamma^{*\top} \hat \gamma^* + \hat \beta^2) - 1} \cdot \frac{\exp(\hat \gamma^\top \hat \gamma)}{\exp(\hat \gamma^{*\top} \hat \gamma^* + \hat \beta^2)}.$$
}
}
\begin{proof} 
Because $[\bX_i, U_i] \iid MVN(0, I)$, $\hat w_i$ and $\hat w_i^*$ both are lognormal random variables, by definition, the variance of $\hat w_i^*$ is: $(\exp(\hat \gamma^{*\top} \gamma^* + \hat \beta^2) - 1)\cdot \exp(\hat \gamma^{*\top} \hat \gamma + \hat \beta^2),$ and similarly, the variance of $\hat w_i$ is: $(\exp(\hat \gamma^\top \hat \gamma) - 1)\cdot \exp(\hat \gamma^\top \hat \gamma)$. Then, the result of the example immediately follows, using $R^2 := 1 - \var(w_i \mid A_i = 1)/\var(w_i^* \mid A_i = 1).$

\end{proof} 
\subsection{Corollary \ref{cor:asy_vsm}}
\noindent\fbox{%
\vspace{2mm}
\parbox{\textwidth}{%
\vspace{2mm}
Consider the set of confounders, in which for all $\delta > 0$, $P(w^*_i / w_i < \delta) > 0$, or $P(w^*_i / w_i > \delta) > 0$. Then, if the outcomes are unbounded, the threshold from Theorem \ref{thm:bound_cond} will converge in probability to 1:
$$\frac{\psi(\Lambda)^2}{4 (1-\cor(w_i, Y_i)^2) \cdot \var(w_i) \var(Y_i) + \psi(\Lambda)^2} \cip 1$$
}
}
\begin{proof} 
To begin, for simplicity of notation, we define $g(\psi(\Lambda); Y_i, w_i)$ as the threshold from Theorem \ref{thm:bound_cond}:
$$g(\psi(\Lambda); Y_i, w_i) := \frac{\psi(\Lambda)^2}{4 (1-\cor(w_i, Y_i)^2) \cdot \var(w_i) \var(Y_i) + \psi(\Lambda)^2}$$

We will use results from \cite{resnick2008extreme}, who show that for a sequence of random variables drawn i.i.d., the maximum of the sequence will converge in probability towards the upper bound of the support. We provide the derivation for completeness. Let $\{W_i\}_{i=1}^n$ be drawn i.i.d. from a distribution $F$. Then by i.i.d.: 
$$P(W^{(n)} \leq w) =  P \left(\bigcap_{i=1}^n \{W_i \leq w\} \right)= F_Y^{n}(w),$$
where $W^{(1)} \leq ... \leq W^{(n)}$. Then define $w_0 = \sup\{w:F_W(w) < 1\}$. Then for some $w' < w_0$, $P(W^{(n)} \leq w') = F_W^n(w') = 0$, since $F_W(w') < 1$. As such, $W^{(n)}$ converges almost surely, and by extension, in probability, to $w_0$. We note that the same result can be applied for the minima of $\{W_i\}_{i=1}^n$ by using $-\{W_i\}_{i=1}^n$. 

Now, define $\lambda_i := w_i^*/w_i$. We have restricted the set of plausible $\lambda_i$ such that $\lim\inf\{\lambda: 1-F_\lambda(\lambda) <1\} = 0$, or $\lim\sup\{\lambda:F_\lambda(\lambda) < 1\} \to \infty$. First consider the setting for $\lim\inf\{\lambda: 1-F_\lambda(\lambda) <1\} = 0$. We can apply the results from above to show that for a sequence of random $\lambda_1, ..., \lambda_n$, the minimum of the sequence will converge in probability towards zero Because $\Lambda = \max_{1 \leq i \leq n} \left\{\lambda_i, 1/\lambda_i\right \}$, this implies that $\Lambda$ will diverge in probability towards infinity. Similarly, for $\lim\sup\{\lambda:F_\lambda(\lambda) < 1\} \to \infty$, the maximum of the sequence will diverge in probability towards infinity, which implies that $\Lambda$ will diverge in probability towards infinity. 

Applying Continuous Mapping Theorem and sample boundedness, the length of the point estimate bounds under the marginal sensitivity model ($\psi(\Lambda)$) will be equal to the range of the observed control outcomes. Thus, if the outcomes $Y_i$ are unbounded as well (i.e., $F_Y(y) < 1$ for all $y \in \R$.), $\psi(\Lambda)$ will diverge in probability to infinity. 

Thus, we have shown that $\psi(\Lambda)$ will diverge in probability to infinity. Applying Continuous Mapping Theorem again, $g(\psi(\Lambda); Y_i, w_i) \cip 1$, which concludes the proof. 

\end{proof}

\section{Extended Tables}\label{app:extra_tables}
\begin{table}[ht]
\centering
\textbf{Benchmarking Results}\\ \vspace{2mm} 
\begin{tabular}{lcc|cccc}
  \toprule
Covariate & $\hat \Lambda$ & MSM & $\hat R^2$ & $\hat \rho$ & VBM & VBM, w/ Corr. \\ 
  \midrule
Gender & 1.1& [1.71, 2.57] & 0.00 & 0.01& [1.81, 2.48] & [1.81, 2.48] \\ 
  Age & 2.1 & [0.98, 3.10] & 0.12 & -0.01& [1.24, 3.02] & [1.87, 2.43] \\ 
  Income & 2.9 & [0.56, 3.32] & 0.14 & -0.09 & [1.18, 3.07] & [1.87, 2.43] \\ 
  Income (Missing) & 1.2 & [1.62, 2.64] & 0.00 & -0.05&[1.81, 2.48] & [1.81, 2.48] \\ 
  Education & 4.7 & [-0.06, 3.55] & 0.17 & 0.06& [1.10, 3.15] & [1.77, 2.52] \\ 
  Cig. Smoked & 2.2 &  [0.92, 3.13] &0.01&  -0.01& [1.66, 2.62] & [1.81, 2.48] \\ 
  Smoking History & 1.5 & [1.38, 2.83] & 0.04& -0.02&[1.50, 2.77] & [1.82, 2.47] \\ 
  Race & 3.5 & [0.31, 3.43] & 0.19 & -0.09& [1.05, 3.20] & [1.88, 2.42] \\ 
   \bottomrule
\end{tabular}
\caption{Benchmarking results for both the marginal sensitivity model and the variance-based sensitivity model. We include both the benchmarked parameter values, as well as the estimated 95\% confidence intervals.} 
\end{table}

\end{document}